\documentclass{article}
\usepackage{amsmath,amssymb,theorem}
\usepackage[left=2.5cm, right=3.5cm]{geometry}
\newcommand{\TeXmacs}{T\kern-.1667em\lower.5ex\hbox{E}\kern-.125emX\kern-.1em\lower.5ex\hbox{\textsc{m\kern-.05ema\kern-.125emc\kern-.05ems}}}
\newcommand{\assign}{:=}
\newcommand{\asterisk}{*}
\newcommand{\bignone}{}
\newcommand{\tmem}[1]{{\em #1\/}}
\newcommand{\tmmathbf}[1]{\ensuremath{\boldsymbol{#1}}}
\newcommand{\tmop}[1]{\ensuremath{\operatorname{#1}}}
\newcommand{\tmtextit}[1]{{\itshape{#1}}}
\newcommand{\withTeXmacstext}{This document has been produced using \TeXmacs (see \texttt{http://www.texmacs.org})}
{\theorembodyfont{\rmfamily}\newtheorem{remark}{Remark}}
\newtheorem{theorem}{Theorem}
\begin{document}

\title{Spin Matrix for the Scaled Periodic Ising
Model\thanks{{\withTeXmacstext}}}\author{John Palmer and Grethe
Hystad}\maketitle

In this paper we consider the matrix representation for the spin operator in
the continuum limit of the periodic Ising model. \ We fix the interval $[- L,
L]$ as the target for the continuum limit in the $x$ variable with periodic
boundary conditions $\sigma (L) = \sigma (- L)$ for the spin operator. \ The
scaling limit we are interested in is a limit in which the lattice spacing,
$\delta_T,$ for the Ising model in the interval $[- L, L]$ tends to 0 as the
temperature $T \uparrow T_c$ approaches the critical temperature. The
relationship between the lattice spacing and the temperature is determined by
taking the inverse, $\delta_T^{- 1}$, to be the correlation length at
temperature $T$. \ The interval $[- L, L]$ will then have length $2 L$ in
units of correlation length throughout the limiting process. \ The results for
the matrix elements of the Ising spin operator will have consequences for the
correlations in the continuum limit for the cylinder $[- L, L] \times
\tmmathbf{R}$ and also for the torus $[- L, L] \times [- M, M]$. \

We will not attempt to control the convergence of the scaling limit in this
paper.  This is partly because the convergence issue is much simplified if
one uses formulas for the matrix elements on a finite periodic lattice that
were conjectured by Bugrij and Lisovyy (see {\cite{B01}}, {\cite{BL03}} and
{\cite{BL04}}). A proof of these formulas has appeared but is quite
complicated {\cite{GIPST07}} and {\cite{GIPST07a}}. \ We expect that the
technique we use in this paper can be extended to deal with the finite
periodic lattice to give an alternative proof of the Bugrij-Lisovyy
conjecture. The results for the continuum limit that we address in this paper
were already announced in {\cite{FZ03}} and we should mention that the
principal technique we employ is a Green function construction that we learned
from the paper of Lisovyy {\cite{L05}}.

We begin by recounting some results from Grethe Hystad's thesis {\cite{GH09}},
that provide a representation for the continuum limit of the Ising model
(under the hypothesis that the Bugrij-Lisovyy conjecture is correct). The
framework for her thesis is a reworking of Bruria Kaufmann's 1948 paper on the
periodic Ising model {\cite{K49}} We can avoid some extraneous detail if we
limit our considerations at the start to a finite lattice $\Lambda_{\ell, m}
=\{- \ell, \cdots, \ell\} \times \{- m, \cdots, m\}$ where $\ell$ and $m$ are
positive integers. \ A configuration of spins on the lattice is a map,
\[ \sigma : \Lambda_{\ell, m} \rightarrow \{- 1, 1\}. \]
In this paper we are exclusively interested in the periodic boundary
conditions $\sigma (- \ell, j) = \sigma (\ell, j)$ for all $j.$ We will be
interested both in the boundary conditions for periodic behavior in the
vertical direction, $\sigma (j, - m) = \sigma (j, m)$ for all $j$ and also in
the cylindrical limit $m \rightarrow \infty$. \ The reader might note that
little appears to be gained in the periodic situation by having the lattice
sites run from $- \ell$ to $\ell$ rather than from 0 to $\ell$ as is perhaps
customary. \ However, there are some differences in our treatment of the model
for odd and even numbers of horizontal lattice sites that make it simpler to
confine our attention to the odd case. \ \ The energy of a periodic
configuration for the Ising model is,
\[ E_{\ell, m} (\sigma) = - \sum_{\langle i, j \rangle} J_{i, j} \sigma (i)
   \sigma (j) \]
where the sum is over nearest neighbors $i$ and $j$ in $\Lambda_{\ell, m}$ and
in our considerations there are only two interaction strengths, $J_{i, j} =
J_1 > 0$ when $i$ and $j$ are horizontal neighbors and $J_{i, j} = J_2 > 0$
when $i$ and $j$ are vertical neighbors. \ Of course, in the periodic case
lattice points with the same second coordinate and first coordinates $\ell$
and $- \ell$ are nearest neighbors as are points with the same first
coordinate and second coordinates $m$ and $- m.$ \ The partition function at
temperature $T$ is the sum of the Boltzmann weights associated to a
configuration,
\[ Z_{\ell, m} = \sum_{\sigma} \exp (- E_{\ell, m} (\sigma) / k_B T) . \]
The Boltzmann constant, $k_B$, appears in this formula, for reasons of
tradition, but nothing is lost for us replacing $J_j$ by $J_j / k_B$ and
setting $k_B = 1$.

Kaufmann's basic result is a formula for this partition function as the trace
of the $2 m + 1$ power of a transfer matrix, $V,$ that has a characterization
in the spin representation of the orthogonal group,
\[ Z_{\ell, m} = \tmop{Tr} (V^{2 m + 1}) . \]

\section{Transfer Matrix}

Our first goal is to summarize the reformulation of Kaufmann's result for $V$
that can be found in the dissertation {\cite{GH09}}. \ The transfer matrix $V$
acts on the tensor product,
\[ \mathcal{H} = \bigotimes_{n = - \ell}^{\ell} \tmmathbf{C}^2_n, \]
where $\tmmathbf{C}^2_n$ is just a copy of $\tmmathbf{C}^2$. \ Suppose that
$X$ is a map on $\tmmathbf{C}^2$. \ Let $X_n$ denote the linear transformation
on $\mathcal{H}$ that acts as $X$ on the $n^{\tmop{th}} $ slot and the
identity in the remaining slots. There is a representation of the Clifford
algebra on $\mathcal{H} $ determined by the action of generators,
\begin{eqnarray}
  q_n = \left( \prod^{n - 1}_{k = - \ell} X_k \right) Y_n, &  &  \nonumber\\
  p_n = \left( \prod_{k = - \ell}^{n - 1} X_k \right) Z_n, &  & \label{eq1} 
\end{eqnarray}
where,

\[ X = \left(\begin{array}{cc}
     0 & 1\\
     1 & 0
   \end{array}\right), Y = \left(\begin{array}{cc}
     0 & - i\\
     i & 0
   \end{array}\right), Z = \left(\begin{array}{cc}
     1 & 0\\
     0 & - 1
   \end{array}\right) . \]
Then $\{q_k, p_k \}$ satisfy the usual generator relations for the Clifford
algebra,
\begin{eqnarray*}
  &  & p_k p_l + p_l p_k = 2 \delta_{k l}, q_k q_l + q_l q_k = 2 \delta_{l
  k}, q_k p_l + p_l q_k = 0.
\end{eqnarray*}
We introduce the vector space $W$ of complex linear combinations of $q_k$ and
$p_k$ with coordinates,
\[ W \ni w = \frac{1}{\sqrt{2}} \sum^{\ell}_{k = - \ell} x_k (w) q_k + y_k (w)
   p_k, \]
and distinguished complex bilinear form,
\[ (w, w') = \sum_{k = - \ell}^{\ell} x_k (w) x_k (w') + y_k (w) y_k (w') . \]
The $\{q_k, p_k \}$ are then generators of an irreducible representation of
the Clifford algebra $\tmop{Cliff} (W)${\cite{BW35}}. \ The conjugation,
\[ \overline{w} = \frac{1}{\sqrt{2}} \sum^{\ell}_{k = - \ell}
   \overline{x_{}}_k q_k + \overline{y}_k p_k, \]
determines an Hermitian inner product on $W,$
\[ \langle w, w' \rangle = ( \overline{w}, w') . \]
The vectors $q_k, p_k$ are real with respect to this conjugation and since
they are self-adjoint with respect to the standard inner product on the tensor
product of copies of the Hermitian inner product space $\tmmathbf{C}^2$ this
representation of $\tmop{Cliff} (W)$ is a $\asterisk - \tmop{representation}$.

An important role in Kaufmann's analysis is played by the parity operator,
\[ U = \prod_k X_k = \prod_k i p_k q_k . \]
Evidently $U^2 = 1$ and for reasons that will be apparent shortly we write
$\mathcal{H_{}}_A$ for the $+ 1$ eigenspace of $U$ and $\mathcal{H}_P$ for the
$- 1$ eigenspace of $U$. \ The transfer matrix $V$ respects the direct sum
decomposition $\mathcal{H} = \mathcal{H}_A \oplus \mathcal{H}_P$ and we write,
\[ V = V_A \oplus V_P . \]
In order to characterize the maps $V_A$ and $V_P$ we introduce the finite
Fourier transforms $\mathcal{F}_A$ and $\mathcal{F}_P$. \ Let,
\[ \mathcal{I}_{\ell} =\{- \ell, - \ell + 1, \ldots, \ell\}, \]
and note that in the $\{q_k, p_k \}$ basis we can think of $W$ as the finite
sequence space,
\[ W = \ell^2 ( \mathcal{I}_{\ell}, \tmmathbf{C}^2) . \]
For $f \in \ell^2 ( \mathcal{I}_{\ell}, \tmmathbf{C}^2)$ and $z \in
\tmmathbf{C}$ write,
\[ \mathcal{F} f (z) = \frac{1}{\sqrt{2 \ell + 1}} \sum_{k = - \ell}^{\ell} f
   (k) z^k \]
Let $\Sigma_A =\{z \in \tmmathbf{C}| z^{2 \ell + 1} = - 1\}$ and $\Sigma_P
=\{z \in \tmmathbf{C}| z^{2 \ell + 1} = 1\}$. We refer to $\Sigma_A$ and
$\Sigma_P$ as the anti-periodic and periodic spectrum on the unit circle. \
Then $\mathcal{F}_A f$ and $\mathcal{F}_P f$ are respectively defined as
the restrictions of $\mathcal{F} f$ to the anti-periodic and periodic spectral
points on the unit circle,
\begin{eqnarray}
  \mathcal{F}_A f (z) = \mathcal{F} f (z) \tmop{for} z \in \Sigma_A, &  & 
  \nonumber\\
  \mathcal{F}_P f (z) = \mathcal{F} f (z) \tmop{for} z \in \Sigma_P
  . \label{fourier} &  & 
\end{eqnarray}
It is easy to confirm the inversion formulas,
\begin{eqnarray*}
  f (k) = \frac{1}{\sqrt{2 \ell + 1}} \sum_{z \in \Sigma_A} \mathcal{F}_A f
  (z) z^{- k} \tmop{for} k \in \mathcal{I}_{\ell}, &  & \\
  f (k) = \frac{1}{\sqrt{2 \ell + 1}} \sum_{z \in \Sigma_P} \mathcal{F}_P f
  (z) z^{- k} \tmop{for} k \in \mathcal{I}_{\ell} . &  & 
\end{eqnarray*}
For $|z| = 1$ define a $2 \times 2$ matrix,
\[ T_z (V) = e^{- \gamma (z)} Q_+ (z) + e^{\gamma (z)} Q_- (z), \tmop{with}
   Q_{\pm} (z) = \frac{1}{2} \left(\begin{array}{cc}
     1 & \mp w (z)\\
     \mp \overline{w (z)} & 1
   \end{array}\right) \]
where $\gamma (z)$ and $w (z)$ are defined by,
\begin{eqnarray*}
  \tmop{ch} \gamma (z) = & c_2^{\asterisk} c_1 - s_2^{\asterisk} s_1 (z + z^{-
  1}) / 2 & 
\end{eqnarray*}
\[ w (z) = i \frac{\mathcal{A}_1 (z) \mathcal{A}_2 (z)}{\mathcal{A}_1
   (z^{- 1}) \mathcal{A}_2 (z^{- 1})} . \]
Note that for brevity we write,
\[ \tmop{ch} x = \cosh x, \tmop{and} \tmop{sh} x = \sinh x. \]
The constants $s_j \tmop{and} c_j$ and their ``duals'' $s_j^{\asterisk}
\tmop{and} c_j^{\asterisk}$ are defined by,
\[ s_j = \tmop{sh} (2 J_j / k_B T), c_j = \tmop{ch} (2 J_j / k_B T), \]
\[ s_j^{\asterisk} = s_j^{- 1}, c_j^{\asterisk} = c_j s_j^{- 1}, \]
and finally,
\[ \mathcal{A}_j (z) = \sqrt{\alpha_j - z} \]
with,
\[ \alpha_1 = (c_1^{\asterisk} - s_1^{\asterisk}) (c_2 + s_2), \alpha_2 =
   (c_1^{\asterisk} + s_1^{\asterisk}) (c_2 + s_2) \]
It will be simplest for us to confine our attention to what happens when the
temperature, $T$, is strictly less than the critical temperature ($T <
T_c)${\cite{JP06}}. \ In this case $\alpha_2 > \alpha_1 > 1$ and smooth square
roots $S^1 \ni z \rightarrow \mathcal{A}_j (z)$ are uniquely determined by the
normalization, $\mathcal{A}_j (1) > 0 .$

Let $T_A (V)$ denote the operator on $W$ whose action in the $\mathcal{F}_A$
representation is given by,
\[ \mathcal{F}_A f (z) \rightarrow T_z (V) \mathcal{F}_A f (z), \]
with a completely analogous definition for $T_P (V) .$ \ Let $Q_A^{\pm}$
denote the operator on $W$ whose action in the $\mathcal{F}_A$
representation is given by,
\[ \mathcal{F}_A f (z) \rightarrow Q_{\pm} (z) \mathcal{F}_A f (z), \]
with an analogous definition for $Q_P^{\pm}$. \

Both $T_A (V)$ and $T_P (V)$ have positive real spectrum and neither has 1 as
an eigenvalue (for finite $\ell$). \ Let $W_A^+ = Q_A^+ W$ denote the span of
the eigenvectors for $T_A (V)$ that have eigenvalues between 0 and 1. \ and
let $W_A^- = Q_A^- W$ denote the span of the eigenvectors for $T_A (V)$ that
have eigenvalues greater than 1. \ Let $T_A^+$ denote the restriction of $T_A
(V)$ to the subspace $W_A^+$. \ Make precisely analogous definitions for
$W_P^{\pm}$ and $T_P^+$. \

In the dissertation {\cite{GH09}} it is proved that $\mathcal{H}_A$ is
unitarily equivalent to the even tensor algebra over $W_A^+$. That is,
\[ \mathcal{H}_A \simeq \tmop{Alt}_{\tmop{even}} (W_A^+)
   =\tmmathbf{C} \oplus W_A^+ \wedge W_A^+ \oplus \cdots \oplus \Lambda^{2
   \ell W_A^+}, \]
where $\Lambda^k W_A^+$ is the $k$ fold alternating tensor product of $W_A^+$
with itself, and that in this representation,
\begin{equation}
  V_A = \lambda_A (1 \oplus T_A^+ \otimes T_A^+ \oplus \cdots \oplus
  (T_A^+)^{\otimes 2 \ell}), \label{eq5}
\end{equation}
where $\lambda_A$ is the largest eigenvalue of $V_A$ given by,
\[ \lambda_A = \exp \frac{1}{2} \sum_{z \in \Sigma_A} \gamma (z) . \]
In a similar fashion $\mathcal{H}_P$ is unitarily equivalent to the even
tensor algebra over $W_P^+$,
\begin{equation}
  \mathcal{H}_P \simeq \tmop{Alt}_{\tmop{even}} (W_P^+)
  =\tmmathbf{C} \oplus W_P^+ \wedge W_P^+ \oplus \cdots \oplus \Lambda^{2
  \ell} W_P^+, \label{eq6}
\end{equation}
with,
\[ V_P = \lambda_P (1 \oplus T_P^+ \otimes T_P^+ \oplus \cdots \oplus
   (T_P^+)^{\otimes 2 \ell}), \]
where $\lambda_P$ is the largest eigenvalue of $V_P$ given by,
\[ \lambda_P = \exp \frac{1}{2} \sum_{z \in \Sigma_P} \gamma (z) . \]
\begin{remark}
  This representation of $T_z (V)$ differs from the representation in {\cite{GH09}} by
  conjugation by,
  \[ \left(\begin{array}{cc}
       z & 0\\
       0 & 1
     \end{array}\right), \]
\end{remark}

representing translation by 1 lattice unit in the $q_k$ basis elements (with
$q_{\ell + 1}$ equal to$- q_{- \ell}$, or $q_{- \ell} $ depending on which
transform, $\mathcal{F}_A$ or $\mathcal{F}_P$, is relevant).

This is convenient since the corresponding ``conjugation'' acting on the
induced rotation for the spin operator, reduces that operator to the
``difference'' of translations acting in the periodic and anti-periodic
sectors.

\section{Spin operator}

The spin operator at $(j, 0)$, which we write as $\sigma_j$ acts on the
Clifford generators,
\begin{eqnarray}
  \sigma_j q_k \sigma_j^{- 1} = - \tmop{sgn} (k - j - 1) q_k &  & \\
  \sigma_j p_k \sigma_j^{- 1} = - \tmop{sgn} (k - j - 1) p_k & \label{eq2} & 
  \nonumber
\end{eqnarray}
Again, compared to {\cite{GH09}} this is conjugated by
$\left(\begin{array}{cc}
  z & 0\\
  0 & 1
\end{array}\right)$. \ The spin operators $\sigma_j$ anti-commute with $U$
and hence map $\mathcal{H}_A$ into $\mathcal{H}_P $ and vice-versa. \ Thus we
write,
\[ \sigma_j = \left(\begin{array}{cc}
     0 & \sigma_j^{A P}\\
     \sigma_j^{P A} & 0
   \end{array}\right) \tmop{acting} \tmop{on} \mathcal{H}_A \oplus
   \mathcal{H}_P . \]

One way to make use of $( \ref{eq2})$ is to note that $\mathcal{H} =
\mathcal{H}_A \oplus \mathcal{H}_P$ is also unitarily equivalent to
both, $\tmop{Alt} (W_A^+)$ and $\tmop{Alt} (W_P^-)$. \ Thus we can regard
$\sigma_j$ as a map,
\begin{equation}
  \sigma_j : \tmop{Alt} (W_A^+) \rightarrow \tmop{Alt} (W_P^+), \label{eq3}
\end{equation}
each space carrying an irreducible $\asterisk -$representation \ of the
Clifford algebra $\tmop{Cliff} (W)$. The unitary map, $U_A$, that allows us to
(projectively) identify $\mathcal{H} = (\tmmathbf{C}^2)^{\otimes 2 \ell + 1}$
with $\tmop{Alt} (W_A^+)$ intertwines the action of the Clifford algebra on
$\mathcal{H}$ determined by the generators $q_k, p_k$ with the Fock
representation $F_A$ on $\tmop{Alt} (W_A^+) .$ That is, for $w = \sum x_k q_k
+ y_k p_k$ we have,
\begin{equation}
  U_A w = F_A (w) U_A . \label{eq4}
\end{equation}
Recall that the Fock representation is given by,
\[ F_A (w) = c (w_A^+) + a (w_A^-), \tmop{where} w_A^{\pm} = Q_A^{\pm} w, \]
and $c (\cdot)$ and $a (\cdot)$ are creation and annihilation operators
{\cite{JP06}}. \ Since $Q_A^{\pm}$ are self-adjoint with respect to the
Hermitian inner product on $W$, $F_A$ determines a $\asterisk -
\tmop{representation}$ of $\tmop{Cliff} (W)$, and since $W$ is finite
dimensional all such irreducible representations are unitarily equivalent. \
Thus $U_A : \mathcal{H} \rightarrow \tmop{Alt} (W_A^+)$ \ exists and
is determined up to a multiple of absolute value 1 by (\ref{eq4}).

The map $\sigma_j$ is then characterized up to a scalar multiple by the
relations (\ref{eq2}) and we can recover $\sigma_j^{P A}$ by restricting
(\ref{eq3}) to the even subspace, $\tmop{Alt}_{\tmop{even}} (W_A^+) .$ \ This
is important for us since we are interested in understanding the matrix
elements of $\sigma_j$ in a basis of eigenvectors for the transfer matrix $V$
and these are simple only in the spaces, $\tmop{Alt}_{\tmop{even}} (W_{A,
P}^+) .$ \ In a similar fashion we can regard,
\[ \sigma_j : \tmop{Alt} (W_P^+) \rightarrow \tmop{Alt} (W_A^+), \]
and understand $\sigma_j^{A P}$ as the restriction of $\sigma_j$ to the even
subspace $\tmop{Alt}_{\tmop{even}} (W_P^+) .$ The reader should keep in mind,
however, that this map is not to be confused with (\ref{eq3}) -- in fact,
since $\sigma_j^2 = 1$, this map can be identified with $\sigma_j^{- 1}$ for
$\sigma_j$ coming from (\ref{eq3}). \

A fairly dramatic simplification of the induced rotation for $\sigma_j$
occurs for $j = \ell$. \ In this case, (\ref{eq2}) implies that the induced
rotation for $\sigma_{\ell}$ is the identity. \ Thus $\sigma_{\ell}$ is an
intertwining map for the Fock representations $F_A$ and $F_P$. \ That is,
\begin{equation}
  \sigma_{\ell} F_A (w) = F_P (w) \sigma_{\ell .} \label{eq7}
\end{equation}
One can recover the spin operators $\sigma_k$ by conjugating with the
appropriate ``space'' translations acting on,
\[ \tmop{Alt}_{\tmop{even}} (W_A^+) \oplus \tmop{Alt}_{\tmop{even}} (W_P^+) .
\]
Translation by 1 lattice unit is given by,
\[ 1 \oplus (z \otimes z) \oplus \cdots \oplus z^{\otimes 2 \ell} \]
in both even tensor algebras. \ However, since $z \in \Sigma_A$ in the first
and $z \in \Sigma_P$ in the second, the spectrum of space translations is
different in each summand. This is reflected by the induced rotation being $2
\ell + 1$anti-periodic in $W = \ell^2 ( \mathcal{I}_m, \tmmathbf{C}^2)$ in the
first summand and $2 \ell + 1$ periodic in the second summand. \ This
difference produces the signs in (\ref{eq2}). \

\section{Eigenvectors of the Transfer Matrix}

The operators $T_A (V)$ and $T_P (V)$ acting on $W$ are self-adjoint. \ We
introduce an orthonormal basis of eigenvectors for these maps. \ Define,
\[ a (z) = \sqrt{\frac{\mathcal{A}_1 (z) \mathcal{A}_2 (z)}{\mathcal{A}_1 (z^{-
   1}) \mathcal{A}_2 (z^{- 1})}}, \]
which we normalize so that it is positive for $z = 1$ (and as usual for us, $T
< T_c$ so the square roots are all holomorphic functions in a neighborhood of
$z \in S^1$). Then for $z \in \Sigma_A$, $T_A$ has eigenvalues $\exp (- \gamma
(z))$ and $\exp (\gamma (z))$ with associated eigenvectors $e_A^+ (z)$ and
$e_A^- (z)$, given as functions in the $\mathcal{F}_A $ representation of $W$
by,
\begin{eqnarray}
  u \rightarrow e_A^+ (z, u) = \frac{1}{\sqrt{2}} \left(\begin{array}{c}
    a (z)\\
    i a (z)^{- 1}
  \end{array}\right) \delta (z, u), \tmop{for} z, u \in \Sigma_A, &  & 
  \nonumber\\
  u \rightarrow e_A^- (z, u) = \frac{1}{\sqrt{2}} \left(\begin{array}{c}
    a (z)\\
    - i a (z)^{- 1}
  \end{array}\right) \delta (z, u), \tmop{for} z, u \in \Sigma_A . &
  \label{evectors} & 
\end{eqnarray}
Here,
\[ \delta (z, u) = \left\{ \begin{array}{l}
     1 \tmop{if} u = z\\
     0 \tmop{if} u \neq z
   \end{array} \right. \]
Then $\{e_A^+ (z), e_A^- (z)\}_{z \in \Sigma_A}$ is an orthonormal basis for
$W$ with respect the Hermitian inner product and,
\[ (e_A^+ (z), e_A^- (z')) = \delta (z, z'), \]
so $\{e_A^+ (z)\}_{z \in \Sigma_A}$ and $\{e_A^- (z)\}_{z \in \Sigma_A}$ are
dual basis for $W_A^+$ and $W_A^-$ with respect to the complex linear pairing
between these two subspaces. \ We make exactly the same definitions for
$e^{\pm}_P (z)$ except that $z \in \Sigma_P$ in this case.

We are interested in the matrix elements for the spin operator $\sigma =
\sigma_{\ell}$ that connect the eigenvectors for $V$ (all wedge products of
the vectors $e_A^+ (z)$ \ and $e^+_P (z)$). \ So, for example, let
\[ \tmmathbf{z}= (z_1, z_2, \ldots, z_k) \tmop{with} z_i \in \Sigma_A
   \tmop{and} z_i \neq z_j \tmop{for} i \neq j ; \]
and write $\tmmathbf{z} \in \Sigma_A$ in such circumstances.

If $k$ is {\tmem{even}}, the vector $e_A^+ (\tmmathbf{z})$ defined by,
\[ e_A^+ (\tmmathbf{z}) = e_A^+ (z_1) \wedge \cdots \wedge e_A^+ (z_k), \]
is an eigenvector for the transfer matrix $V$ with eigenvalue,
\[ \exp \left( \frac{1}{2} \sum_{z \in \Sigma_A} \gamma (z) - \sum_{z \in
   \tmmathbf{z}} \gamma (z) \right) . \]
We write $z \in \tmmathbf{z}$ iff $z = z_j \tmop{for} \tmop{some} j = 1, 2,
\ldots, k$, and we write $k =\#\tmmathbf{z}$. \ Exchanging the roles of $P$
and $A$ let,
\[ \tmmathbf{z}' = (z'_1, z'_2, \ldots, z'_k) \tmop{with} z'_i \in \Sigma_P
   \tmop{and} z'_i \neq z'_j \tmop{for} i \neq j. \]
Then if $k$ is {\tmem{even}},
\[ e_P^+ (\tmmathbf{z}') = e_P^+ (z'_1) \wedge \cdots \wedge e_P^+ (z'_k) \]
is an eigenvector for the transfer matrix $V$ with eigenvalue,
\[ \exp \left( \frac{1}{2} \sum_{z' \in \Sigma_P} \gamma (z) - \sum_{z' \in
   \tmmathbf{z}'} \gamma (z') \right) . \]

\section{Matrix of the spin operator}

The matrix element,
\[ \langle e_P^+ (\tmmathbf{z}'), \sigma e_A^+ (\tmmathbf{z}) \rangle
   \tmop{for} \tmmathbf{z}' \in \Sigma_P \tmop{and} \tmmathbf{z} \in \Sigma_A,
\]
makes sense if we think of $\sigma$ as a map,
\[ \sigma : \tmop{Alt} (W_A^+) \rightarrow \tmop{Alt} (W_P^+) . \]
Of course, $e_P^+ (\tmmathbf{z}')$ and $e_A^+ (\tmmathbf{z})$ are eigenvectors
for $V$ only if both $\#\tmmathbf{z}'$ and $\#\tmmathbf{z}$ are even. However
it is easier to explain the version of Wick's theorem that allows us to reduce
the calculation of these matrix to a few basic types if we admit matrix
elements for arbitrary wedge products.

Let $0_A$ and $0_P$ denote the unit vacuum vectors in $\tmop{Alt} (W_A^+)$ and
$\tmop{Alt} (W_P^+)$ (the reader might note that these vectors are only
defined up a multiplier of absolute value 1 by the abstract unitary
equivalence of these spaces with the original tensor product -- we will have
more to say about this later). \ The induced rotation $T (\sigma)$ is the
identity on $W$. \ We write,
\[ T (\sigma) = \left(\begin{array}{cc}
     A & B\\
     C & D
   \end{array}\right), \]
for the matrix of this map taking one from the $W_A^+ \oplus W_A^-$ splitting
of $W$ to the $W_P^+ \oplus W_P^-$ splitting. So, for example $D = Q_P^-
|_{W_A^-}$, maps $W_A^-$ into $W_P^-$. \ We are particularly interested in the
case where $D$ is invertible. \ This happens precisely when $\langle 0_P,
\sigma 0_A \rangle \neq$ 0. In this circumstance one has the following
formulas for the simplest matrix elements of $\sigma$ {\cite{GH09}},
\[ \frac{\langle e_P^+ (z'), \sigma e_A^+ (z) \rangle}{\langle 0_P, \sigma 0_A
   \rangle} = (e_P^- (z'), D^{- \tau} e_A^+ (z)) = D^{- \tau}_{z', z} \]
\[ \frac{\langle e_P^+ (z_1') \wedge e_P^+ (z_2'), \sigma 0_A \rangle}{\langle
   0_P, \sigma 0_A \rangle} = (e_P^- (z_1'), B D^{- 1} e_P^- (z_2')) = B D^{-
   1}_{z_1', z_2'} \]
and
\[ \frac{\langle 0_P, \sigma e_A^+ (z_1) \wedge e_A^+ (z_2)}{\langle 0_P,
   \sigma 0_A \rangle} = (e_A^+ (z_1), D^{- 1} C e_A^+ (z_2)) = D^{- 1}
   C_{z_1, z_2} \]
Here, $D^{\tau} : W_P^+ \rightarrow W_A^+$ is the transpose of $D$ with
respect to the bilinear form $(\cdot, \cdot)$ and
\[ D^{- \tau} : W_A^+ \rightarrow W_P^+, \]
is the inverse $(D^{\tau})^{- 1}$.

The following result is an extension of a well known reduction formula for the
matrix elements of an element in the spin representation of the orthogonal
group {\cite{JP06}}.

\begin{theorem}
  \label{th:2}Let $\tmmathbf{z} \in \Sigma_A$ and $\tmmathbf{z}' \in
  \Sigma_P$. Then, supposing $\langle 0_P, \sigma 0_A \rangle \neq 0$,
  \[ \frac{\langle e_P^+ (\tmmathbf{z}'), \sigma e_A^+ (\tmmathbf{z})
     \rangle}{\langle 0_P, \sigma 0_A \rangle} = \tmop{Pf}
     \left(\begin{array}{cc}
       R_{\tmmathbf{z}' \times \tmmathbf{z}'} & R_{\tmmathbf{z}' \times
       \tmmathbf{z}}\\
       R_{\tmmathbf{z} \times \tmmathbf{z}'} & R_{\tmmathbf{z} \times
       \tmmathbf{z}}
     \end{array}\right), \]
  where $\tmop{Pf}$ is the Pfaffian and the skew symmetric matrix R has matrix
  elements,
  \[ \left( R_{\tmmathbf{z}' \times \tmmathbf{z}'} \right)_{i, j} =
     \frac{\langle e_P^+ (z_i') \wedge e_P^+ (z_j'), \sigma 0_A
     \rangle}{\langle 0_P, \sigma 0_A \rangle} = B D^{- 1}_{z_i', z_j'} \]
  
  \[ \left( R_{\tmmathbf{z}' \times \tmmathbf{z}} \right)_{i, j} = - \left(
     R_{\tmmathbf{z} \times \tmmathbf{z}'} \right)_{j, i} = \frac{\langle
     e_P^+ (z_i'), \sigma e_A^+ (z_j) \rangle}{\langle 0_P, \sigma 0_A
     \rangle} = D^{- \tau}_{z_i', z_j} \]
  and
  \[ \left( R_{\tmmathbf{z} \times \tmmathbf{z}} \right)_{i, j} =
     \frac{\langle 0_P, \sigma e_A^+ (z_i) \wedge e_A^+ (z_j)}{\langle 0_P,
     \sigma 0_A \rangle} = D^{- 1} C_{z_i, z_j} \]
\end{theorem}

This result is true even if $\#\tmmathbf{z}'$ and $\#\tmmathbf{z}$ are not
even, but the matrix elements for $\sigma$ no longer connect eigenvectors for
the transfer matrix in that case. \ It is evident that to find explicit
formulas for the spin matrix elements by this method one needs to calculate
$D^{- 1}$. \ We turn to the principal result in this paper, a calculation of
$D^{- 1}$ in the periodic scaling limit.

\section{The periodic scaling limit}

As mentioned above the limit we are interested in is a continuum limit for a
lattice theory that ``lives'' on the real domain $[- L, L] \times
\tmmathbf{R}$, with a lattice spacing that tends to 0 as the temperature tends
to the critical temperature. \ Periodic boundary conditions on $[- L, L]$ are
maintained for the spins throughout. For the cylindrical case ($M \rightarrow
\infty)$ the correlations can be understood as $0_A$ vacuum expectations of
products of spin operators {\cite{GH09}}. \ The doubly periodic case the spin
variable lives on the torus $[- L, L] \times [- M, M] / \sim$(the equivalence
identifies $L \sim - L$ and $M \sim - M$ in the first and second factors
respectively) and correlations are expressed as traces of the same products. \
In both cases one obtains explicit formulas that depend only on the formulas
for the matrix elements of the spin operators. We examine what happens to the
transfer matrix and the spin operator in this continuum limit without worrying
too much about convergence questions. \ The thesis {\cite{GH09}} contains
results for the convergence problem assuming the Bugrij-Lisovyy conjecture
{\cite{BL03}}.

The horizontal and vertical correlation lengths are known for the infinite
volume Ising model. \ They are the reciprocals of the horizontal and vertical
masses defined by,
\begin{eqnarray*}
  m_1 (T) = 2 K_2 - 2 K_1^{\asterisk}, &  & \\
  m_2 (T) = 2 K_1 - 2 K_2^{\asterisk}, &  & 
\end{eqnarray*}
where $K_j = J_j / k_B T$, and $K_j^{\asterisk}$ is defined by,
\[ \tmop{sh} (2 K_j) \tmop{sh} (2 K_j^{\asterisk}) = 1. \]
The limit we are interested in has $J_j / k_B$ fixed and $T \uparrow T_c$,
where the critical temperature, $T_c$, is defined by,
\[ \tmop{sh} \left( \frac{2 J_1}{k_B T_c} \right) \tmop{sh} \left( \frac{2
   J_2}{k_B T_c} \right) = 1. \]
The critical temperature is defined by the self-dual condition,
$K^{\asterisk_{}}_1 = K_2$, and one sees that both $m_1 (T)$ and $m_2 (T)$
tend to zero as $T \uparrow T_c$. \ We look at the model with horizontal
lattice spacing $m_1 (T)$ and vertical lattice spacing $m_2 (T)$. \ More
specifically fix $L > 0$ and let $\ell (T) = [L m_1 (T)^{- 1}]$ where $[x]$ is
the greatest integer less than $x$. Consider the finite Fourier transform on,
\[ W = \ell^2 ( \mathcal{I}_{\ell (T)}, \tmmathbf{C}^2), \]
written out in terms of the scaled variable $x \in m_1 (T) \mathcal{I}_{\ell
(T)} \subset [- L, L]$,
\[ \mathcal{F} f (k) = \frac{1}{\sqrt{2 \ell (T) + 1}} \sum_{x_{}
   \in m_1 (T) \mathcal{I}_{\ell (T)}} f (x / m_1 (T)) \exp \left( - \frac{2
   \pi i k}{2 \ell (T) + 1} \frac{x_{}}{m_1 (T)} \right) . \]
The map,
\[ f \rightarrow F (x) = \frac{1}{\sqrt{m_1 (T)}} f (x / m_1 (T)), \]
is a unitary map from the $\ell^2$ space on $\mathcal{I}_{\ell (T)}$ to the
$\ell^2$ sequence space on $^{} m_1 (T) \mathcal{I}_{\ell (T)}$, with points
weighted by the mass $m_1 (T)$. The naive limit of this Fourier transform as
$T \uparrow T_c$ is,
\begin{equation}
  \mathcal{F} F (k) = \frac{1}{\sqrt{2 L}} \int_{- L}^L F (x) \exp \left( -
  \frac{i \pi k x}{L} \right) d x, \label{cfourier}
\end{equation}
where $k \in \tmmathbf{Z}$ in the case of the periodic Fourier transform and
$k \in \tmmathbf{Z}+ 1 / 2$ for the anti-periodic Fourier transform. Note that
we have changed the look of the finite Fourier transform (\ref{fourier}) by
the substitution,
\[ z \leftarrow \exp \left( - \frac{2 \pi i k}{2 \ell + 1} \right)
   \tmop{where} k \in \tmmathbf{Z}(\tmop{periodic}) \tmop{or} k \in
   \tmmathbf{Z}+ 1 / 2 (\tmop{anti} - \tmop{periodic}) . \]
The parametrization by $- k$ rather than $k$ is motivated by the desire to get
the usual shape for the Fourier series coefficients (\ref{cfourier}). \ The
scaled vertical coordinate is $y \in m_2 (T)\tmmathbf{Z}$. \ To understand the
scaling limit we examine the ``infinitesimal'' generator $\gamma Q$ of $T
(V)$. \ We are interested in the limit,
\[ \lim_{T \uparrow T_c} m_2 (T)^{- 1} \gamma \left( e^{- \frac{2 \pi i k}{2
   \ell (T) + 1}} \right) Q \left( e^{- \frac{2 \pi i k}{2 \ell (T) + 1}}
   \right), \]
where $k \in \tmmathbf{Z}$ or $k \in \tmmathbf{Z}+ 1 / 2$ for the periodic or
anti-periodic sectors respectively. In {\cite{JP06}} it is shown that,
\[ \lim_{T \uparrow T_c} m_2 (T)^{- 1} \gamma \left( e^{- \frac{2 \pi i k}{2
   \ell (T) + 1}} \right) = \sqrt{1 + p^2}, \tmop{where} p = \frac{\pi k}{L} .
\]
Recall that,
\[ Q (z) = - \left(\begin{array}{cc}
     0 & w (z)\\
     \overline{w (z)} & 0
   \end{array}\right) \]
with,
\[ w (z) = i \frac{\mathcal{A}_1 (z) \mathcal{A}_2 (z)}{\mathcal{A}_1 (z^{-
   1}) \mathcal{A}_2 (z^{- 1})} . \]
Since $\alpha_2 > 1$ in the limit $T \uparrow T_c$ we see that,
\[ \lim_{T \uparrow T_c} \frac{\mathcal{A}_2 \left( e^{- \frac{2 \pi i k}{2
   \ell (T) + 1}} \right)}{\mathcal{A}_2 \left( e^{\frac{2 \pi i k}{2 \ell (T)
   + 1}} \right)} = 1. \]
However, $\alpha_1 = e^{m_1 (T)}$ tends to 1 as $T \uparrow T_c $ and we have
the asymptotics,
\[ \mathcal{A}_1 \left( e^{- \frac{2 \pi i k}{2 \ell (T) + 1}} \right) \sim
   \sqrt{m_1 (T)} \sqrt{1 + i p}, \tmop{where} p = \frac{\pi k}{L} . \]
Thus,
\[ \lim_{T \uparrow T_c} \frac{\mathcal{A}_1 \left( e^{- \frac{2 \pi i k}{2
   \ell (T) + 1}} \right)}{\mathcal{A}_1 \left( e^{\frac{2 \pi i k}{2 \ell (T)
   + 1}} \right)} = \frac{\sqrt{1 + i p}}{\sqrt{1 - i p}}, \tmop{for} p =
   \frac{\pi k}{L} . \]
The induced rotation for the transfer matrix thus scales to the relation
between infinitesimal vertical translation and infinitesimal horizontal
translation,
\[ \frac{\partial}{\partial x_2} = \left(\begin{array}{cc}
     0 & 1\\
     1 & 0
   \end{array}\right) p + \left(\begin{array}{cc}
     0 & - i\\
     i & 0
   \end{array}\right) = \left(\begin{array}{cc}
     0 & - i\\
     - i & 0
   \end{array}\right) \frac{\partial}{\partial x_1} +
   \left(\begin{array}{cc}
     0 & - i\\
     i & 0
   \end{array}\right) . \]
where we identify $i p$ with $\partial / \partial x_1$ based on
(\ref{cfourier}). \ Consider the differential equation,
\begin{equation}
  \frac{\partial \psi}{\partial x_2} + \begin{array}{c}
    \left(\begin{array}{cc}
      0 & i\\
      i & 0
    \end{array}\right) \frac{\partial \psi}{\partial x_1} +
    \left(\begin{array}{cc}
      0 & i\\
      - i & 0
    \end{array}\right) \psi = 0. \label{itrans}
  \end{array}
\end{equation}
After making the substitution,
\[ \psi \leftarrow \frac{1}{\sqrt{2}} \left(\begin{array}{cc}
     1 & 1\\
     - 1 & 1
   \end{array}\right) \psi, \]
this differential equation becomes,
\begin{equation}
  \left(\begin{array}{cc}
    1 & - 2 \partial\\
    - 2 \bar{\partial} & 1
  \end{array}\right) \psi = 0, \label{dirac}
\end{equation}
where,
\[ \partial = \frac{1}{2} \left( \frac{\partial}{\partial x_1} - i
   \frac{\partial}{\partial x_2} \right), \tmop{and} \bar{\partial} =
   \frac{1}{2} \left( \frac{\partial}{\partial x_1} + i
   \frac{\partial}{\partial x_2} \right) . \]
Equation (\ref{dirac}) is a convenient form for the (Euclidean) Dirac equation
in two dimensions. \ In fact, it is useful to generalize this just slightly. \
We introduce a mass $m > 0$ into the Dirac equation,
\begin{equation}
  \left(\begin{array}{cc}
    m & - 2 \partial\\
    - 2 \bar{\partial} & m
  \end{array}\right) \psi = 0 \label{mdirac}
\end{equation}
This arises naturally if one introduces the scaling variable $m x_1$ instead
of $x_1$. \ The spin correlations depend only on the product $m L$ so $L$ and
$m$ are not really independent parameters. \ However, in the planar case the
short distance behavior of the scaled correlations can be studied by analyzing
the $m \rightarrow 0$ behavior of solutions to (\ref{mdirac}) and we expect
something similar is possible in this case.

In the scaling limit it is natural to parametrize periodic and anti-periodic
spectrum,
\begin{equation}
  \Sigma_P = \left\{ \frac{\pi}{L} n : n \in \tmmathbf{Z} \right\},
  \label{pspec}
\end{equation}
and
\begin{equation}
  \Sigma_A = \left\{ \frac{\pi}{L} \left( n + \frac{1}{2} \right) : n \in
  \tmmathbf{Z} \right\} . \label{aspec}
\end{equation}
Of course, this notation conflicts with the earlier definition of $\Sigma_{P,
A}$ but for the remainder of the paper we work with the scaling limit so the
possibility of confusion is small. \ The scaling limit of the space $W^+_A$
can be identified as the space of solutions, $\psi (x) = \psi (x_1, x_2)$, to
(\ref{mdirac}) that are $2 L$ {\tmem{anti-periodic}} in $x_1$, are defined in
the upper half plane $\Im (x) = x_2 > 0$ with $L^2 [- L, L]$ boundary values
for $x_2 = 0$ and tend to 0 as $x_2 \rightarrow + \infty$. $W_A^-$ is similar
but the solutions are defined for $x_2 < 0$ and tend to zero as $x_2
\rightarrow - \infty$. The limits for $W_P^{\pm}$ are similar but are defined
in terms of $2 L$ {\tmem{periodic}} solutions for (\ref{mdirac}). We
understand $\sigma_L$ as an interwining map,
\[ \sigma_L : \tmop{Alt} (W_A^+) \rightarrow \tmop{Alt} (W_P^+) . \]
This needs some further elaboration since in the scaling limit the Fock
representations, $F_A$ and $F_P$ are not unitarily equivalent. \ Nonetheless,
for the purposes of finding matrix elements of the molified spin operator
$V^{\varepsilon} \sigma_L V^{\varepsilon'}$ (for $\varepsilon, \varepsilon' >
0$) and formulas for the spin correlations it suffices to invert the $D$
matrix element of the identity map from $W_A^+ \oplus W_A^-$ to $W_P^+ \oplus
W_P^-$. \ This is the problem to which we now turn.

\section{The spectral curve for the Dirac equation}

A crucial ingredient in our analysis of the scaling limit is a spectral curve
for the Dirac equation that allows one to analytically continue the
eigenfunctions that span $W_{A, P}^+$ to the eigenfunctions that span $W_{A,
P}^-$.

For $u \in \tmmathbf{C}$ the function
\[ \tmmathbf{C} \ni x \rightarrow e (x, u) \assign e^{- \frac{m}{2} ( \bar{x}
   u + x u^{- 1})} \left(\begin{array}{c}
     1\\
     - u
   \end{array}\right) \]
is an exponential solution to the Dirac equation
\[ \mathcal{D}e = \left(\begin{array}{cc}
     m & - 2 \partial\\
     - 2 \bar{\partial} & m
   \end{array}\right) e = 0, \]
where $x = x_1 + i x_2$ and \ $\partial = \partial_x = \frac{1}{2} (\partial /
\partial x_1 - i \partial / \partial x_2)$. \ It is convenient to make the
substitution $u = - e^{- s}$ and to regard the complex strip $\Im s \in [-
\pi, \pi]$ with the upper and lower edges identified, $s + 2 \pi i \simeq s,$
as the ``spectral curve'' for the Dirac equation,$\mathcal{D} e =
0.$ We write,
\[ \Sigma (\mathcal{D}) = \tmmathbf{C}/ 2 \pi i\tmmathbf{Z}, \]
for this periodic strip.

For this parametrization the exponential solutions,
\[ E (x, s) = e (x, - e^{- s}) = e^{\frac{m}{2} ( \bar{x} e^{- s} + x e^s)}
   \left(\begin{array}{c}
     1\\
     e^{- s}
   \end{array}\right) \]
that are purely oscillatory in the $x_1$ variable arise for $\Im s = \pm
\frac{i \pi}{2}$. This fact is, in part, the reason we make the choice $u = -
e^{- s} .$ Define
\[ \mathcal{M}_{\pm} = \left\{ s \in \tmmathbf{C}: \Im s = \pm \frac{i \pi}{2}
   \right\} . \]
Fourier analysis allows us to synthesize the solutions to the Dirac equation
of interest to us as linear combinations of exponential solutions, $E (x, s)$,
for $s \in \mathcal{M}_{\pm}$. Write
\[ s_{\pm} = s \pm \frac{i \pi}{2} \]
Then it is easy to check that,
\[ E (x, s_{\pm}) = \exp (m \tmop{ch} (s_{\pm}) x_1 + i m \tmop{sh} (s_{\pm})
   x_2) \left(\begin{array}{c}
     1\\
     e^{- s_{_{\pm}}}
   \end{array}\right) = \exp (\pm i m \tmop{sh} (s) x_1 \mp m \tmop{ch} (s)
   x_2) \left(\begin{array}{c}
     1\\
     \mp i e^{- s}
   \end{array}\right) \]
For $s$ real we see that $E (x, s_+)$ is exponentially small at infinity for
$x_2 > 0$ and $E (x, s_-)$ is exponentially small at infinity for $x_2 < 0$. \
It is useful to record a related result for the norm of $E (x, s)$. \ Let $x =
x_1 + i x_2$ and $s = u + i v$ denote the splitting of $x, \tmop{and} s$ into
real and imaginary parts. \ Then after a short calculation one finds,
\begin{equation}
  ||E (x, s) || = \sqrt{1 + e^{- 2 u}} \exp (m x_1 \tmop{ch} (u) \cos (v) - m
  x_2 \tmop{ch} (u) \sin (v)) \label{Eest} .
\end{equation}

\section{Periodic and anti-periodic Green functions}

Next we calculate the Green function for the operator $\mathcal{D}$ acting on
the smooth $\tmmathbf{C}^2$ valued functions on,
\[ \text{$[- L, L] \times \tmmathbf{R}$, } \]
with either periodic or antiperiodic boundary conditions on $[- L, L]$. \
Introduce the Fourier coefficients,
\[ \hat{f} (p, \xi) = \frac{1}{4 \pi L} \int_{- L}^L d x_1 \int_{-
   \infty}^{\infty} d x_2 \bignone f (x_1, x_2) e^{- i p x_1} e^{- i \xi x_2}
   . \]
The inversion formula for smooth periodic functions in $x_1$ which are small
at $\infty$ in $x_2$ is,
\[ f (x_1, x_2) = \sum_{p \in \Sigma_P} \int_{- \infty}^{\infty} d \xi
   \bignone \hat{f} (p, \xi) e^{i p x_1} e^{i \xi x_2} \bignone \]
The inversion formula for smooth anti-periodic functions in $x_1$ which are
small at $\infty$ in $x_2$ is,
\[ f (x_1, x_2) = \sum_{p \in \Sigma_A} \int_{- \infty}^{\infty} d \xi
   \bignone \hat{f} (p, \xi) e^{i p x_1} e^{i \xi x_2} \bignone . \]
The Green function for $\mathcal{D}$ with a domain that contains the smooth
periodic functions of $x_1 \in [- L, L]$ we denote by $G_P$. The Green
function for $\mathcal{D}$ with a domain that contains the smooth
anti-periodic functions of $x_1 \in [- L, L]$ we denote by $G_A$. For $x = x_1
+ i x_2$ one finds,
\[ G_{P, A} (x) = \frac{1}{4 \pi L} \sum_{p \in \Sigma_{P, A}} \int_{-
   \infty}^{\infty} d \xi \bignone \frac{1}{m^2 + p^2 + \xi^2}
   \left(\begin{array}{cc}
     m & i p + \xi\\
     i p - \xi & m
   \end{array}\right) e^{i p x_1 + i \xi x_2} \bignone . \]
If $x_2 > 0$ the $\xi$ integration can be ``closed'' in the upper half plane
and if $x_2 < 0$ the $\xi$ integration can be ``closed'' in the lower half
plane. \ One finds,
\begin{equation}
  G_{P, A} (x) = \frac{1}{4 L} \sum_{p \in \Sigma_{P, A}} \frac{1}{\omega (p)}
  \left(\begin{array}{cc}
    m & i (p + \omega (p))\\
    i (p - \omega (p)) & m
  \end{array}\right) e^{i p x_1 - \omega (p) x_2} \bignone \tmop{for} x_2 >
  0, \label{eq:1}
\end{equation}
and
\begin{equation}
  G_{P, A} (x) = \frac{1}{4 L} \sum_{p \in \Sigma_{P, A}} \frac{1}{\omega (p)}
  \left(\begin{array}{cc}
    m & i (p - \omega (p))\\
    i (p + \omega (p)) & m
  \end{array}\right) e^{i p x_1 + \omega (p) x_2} \bignone \tmop{for} x_2 <
  0. \label{eq:2}
\end{equation}

To simplify some calculations with these Green functions we introduce,
\[ g_{P, A} (x) = G_{P, A} (x) \left(\begin{array}{cc}
     0 & i\\
     - i & 0
   \end{array}\right) . \]
The functions $g_{P, A} (x - x')$ have the advantage that their columns are
solutions to the homogeneous Dirac equation as a function of $x$ and their
rows are solutions to the homogeneous Dirac equation as functions of $x'$ (for
$G_{P, A}$ one would need to introduce homogeneous solutions to the transpose
of the Dirac equation to describe the rows).

Make the substitution $p = m \tmop{sh} s$ in equation (\ref{eq:1}) and
multiply on the right by $\left(\begin{array}{cc}
  0 & i\\
  - i & 0
\end{array}\right)$. \ One finds,
\begin{equation}
  g_{P, A} (x) = \frac{1}{4 L} \sum_{s \in \Sigma_{P, A}'} \frac{1}{\tmop{ch}
  s} \left(\begin{array}{cc}
    e^s & i\\
    - i & e^{- s}
  \end{array}\right) e^{i m \tmop{sh} (s) x_1 - m \tmop{ch} (s) x_2} \bignone
  \tmop{for} x_2 > 0, \label{eq:3}
\end{equation}
where we've written,
\begin{equation}
  \text{ $s \in \Sigma'_{P, A}$ for $m \tmop{sh} s \in \Sigma_{P, A}$.}
  \label{specprime}
\end{equation}
Make the substitution $p = - m \tmop{sh} s$ in equation (\ref{eq:2}) and
multiply on the right by $\left(\begin{array}{cc}
  0 & i\\
  - i & 0
\end{array}\right)$. One finds
\begin{equation}
  g_{P, A} (x) = \frac{1}{4 L} \sum_{s \in \Sigma_{P, A}'} \frac{1}{\tmop{ch}
  s} \left(\begin{array}{cc}
    - e^s & i\\
    - i & - e^{- s}
  \end{array}\right) e^{- i m \tmop{sh} (s) x_1 + m \tmop{ch} (s) x_2}
  \bignone \tmop{for} x_2 < 0 \label{eq:4}
\end{equation}
Using (\ref{eq:3}), (\ref{eq:4}) and the definition of $E (x, s_{\pm})$ above,
one can verify,
\begin{equation}
  g_{P, A} (x - x') = \frac{1}{4 L} \sum_{s \in \Sigma_{P, A}'} \bignone
  \frac{e^s}{\tmop{ch} s} E (x, s_+) E (x', s_-)^T \tmop{for} \Im (x - x') >
  0, \label{eq:5}
\end{equation}
and
\begin{equation}
  g_{P, A} (x - x') = - \frac{1}{4 L} \sum_{s \in \Sigma_{P, A}'}
  \frac{e^s}{\tmop{ch} s} E (x, s_-) E (x', s_+)^T \bignone \tmop{for} \Im (x
  - x') < 0, \label{eq:6}
\end{equation}
where $X^T$ is the transpose of $X$. \ These formulas will be useful when we
turn to our principal concern--a formula for the Green function of the Dirac
operator on the strip $[- L, L] \times \tmmathbf{R}$with periodic boundary
conditions in the lower half plane ($x_2 < 0$) and anti-periodic boundary
conditions in the upper half plane.

We introduce periodic and anti-periodic spectral transforms that are closely
related to the representations (\ref{eq:5}) and (\ref{eq:6}). \ For $f (x)$ a
smooth periodic ($P$) or smooth anti-periodic ($A$) function of $x \in [- L,
L]$, define spectral transforms $\mathcal{S}_P^{\pm}$ and
$\mathcal{S}_A^{\pm}$
\begin{eqnarray}
  \mathcal{S}_{A, P}^+ f (s) = \frac{1}{2 L} \int_{- L}^L e^{s / 2} E (x,
  s_-)^T f (x) d x \bignone & \tmop{for} s \in \Sigma_{A, P}', &  \nonumber\\
  \mathcal{S}_{A, P}^- f (s) = \frac{1}{2 L} \int_{- L}^L e^{s / 2} E (x,
  s_+)^T f (x) d x \bignone & \tmop{for} s \in \Sigma_{A, P}' . & 
  \label{spec}
\end{eqnarray}

The Fourier inversion formula implies that,
\begin{eqnarray}
  f (x) = \sum_{s \in \Sigma_P'} \frac{e^{s / 2}}{2 \tmop{ch} (s)} \bignone
  \left( \mathcal{S}_P^+ f (s) E (x, s_+) + \mathcal{S}_P^- f (s) E (x, s_-)
  \right) &  &  \nonumber\\
  f (x) = \sum_{s \in \Sigma_A'} \frac{e^{s / 2}}{2 \tmop{ch} (s)} \bignone
  \left( \mathcal{S}_A^+ f (s) E (x, s_+) + \mathcal{S}_A^- f (s) E (x, s_-)
  \right) &  &  \label{inspec}
\end{eqnarray}
Where $f (x)$ is $2 L$ periodic in the first case, and $2 L$ anti-periodic in
the second case (the formulas are true in an appropriate $L^2, \ell^2$ sense
even if the function $f (x)$ is neither periodic or anti-periodic but the
convergence of the inversion sum is especially good if say the function $f$ is
smooth and (anti-)periodic and the spectral transform is the corresponding
(anti-)periodic transform). \ The formulas (\ref{inspec}) respectively
determine the splitting of $f$ in the $W_P^+ \oplus W_P^-$ and $W_A^+ \oplus
W_A^-$ representations of $W = L^2 [- L, L]$. \ The scaling limit analogues of
the projections $Q_{A, P}^{\pm}$ are,
\[ Q_{A, P}^{\pm} f (x) = \sum_{s \in \Sigma_{A.P}'} \bignone \frac{e^{s /
   2}}{2 \tmop{ch} s} \mathcal{S}_{A, P}^{\pm} f (s) E (x, s_{\pm}) . \]
Again, since in this part of the paper we are exclusively concerned with the
continuum limit there should be little danger of a confusion with the earlier
usage for $Q_{A, P}^{\pm}$.

The Plancherel theorem implies,
\begin{eqnarray}
  \frac{1}{2 L} \int_{- L}^L |f (x) |^2 \bignone d x = \sum_{s \in \Sigma_P'}
  \frac{1}{2 \tmop{ch} (s)} \left( | \mathcal{S}_P^+ f (s) |^2 + |
  \mathcal{S}_P^- f (s) |^2 \right) \bignone &  &  \nonumber\\
  \frac{1}{2 L} \int_{- L}^L |f (x) |^2 \bignone d x = \sum_{s \in \Sigma_A'}
  \frac{1}{2 \tmop{ch} (s)} \left( | \mathcal{S}_A^+ f (s) |^2 + |
  \mathcal{S}_A^- f (s) |^2 \right) \bignone &  &  \label{planch}
\end{eqnarray}
It is useful to introduce the components of the inverse spectral transform,
$\mathcal{R},$ as maps,
\begin{eqnarray}
  \mathcal{R}_A^{\pm} f (x) = \sum_{s \in \Sigma_A'} \frac{e^{s / 2}}{2
  \tmop{ch} (s)} f (s) E (x, s_{\pm}) \bignone, &  &  \nonumber\\
  \mathcal{R}_P^{\pm} f (x) = \sum_{s \in \Sigma_P'} \frac{e^{s / 2}}{2
  \tmop{ch} (s)} f (s) E (x, s_{\pm}), \bignone &  &  \label{inspec1}
\end{eqnarray}
where $f \in \ell^2 (\Sigma_A')$ or $f \in \ell^2 (\Sigma_P')$ respectively. \
The Plancherel theorem implies that the maps $\mathcal{R}_{A, P}^{\pm}$ are
all isometries from the appropriate square summable sequence space (with
weight $(2 \tmop{ch} s)^{- 1}$) into $L^2 [- L, L]$.

In preparation for the construction of a Green function for the Dirac operator
that mixes periodic and anti-periodic boundary conditions we follow Lisovyy
{\cite{L05}} and turn next to the solution of a factorization problem on the
spectral curve.

\section{Factorization of $\sigma_P / \sigma_A$ on the spectral curve}

The function,
\[ \sigma_P (s) = (e^{m L \tmop{ch} s} - e^{- m L \tmop{ch} s}) / 2 \]
has simple 0's at the points $s = t \pm i \pi / 2 \in \mathcal{M}_{\pm}$ with
\[ t \in \Sigma_P' . \]
The function,
\[ \sigma_A (s) = (e^{m L \tmop{ch} s} + e^{- m L \tmop{ch} s}) / 2 \]
has simple 0's at the points $s = t \pm i \pi / 2 \in \mathcal{M}_{\pm}$ with
\[ t \in \Sigma'_A . \]
Their reciprocals have simple poles that can be used to express sums over the
points in the periodic and anti-periodic spectrum as contour integrals. \
Lisovyy's construction of the Green function of interest to us depends on the
solution of a factorization problem that we describe next. \ Consider the
functions,  
\[ f_0 (s) = \frac{\sigma_P (s)}{\sigma_A (s)} \tmop{defined} \tmop{for} -
   \epsilon < \Im (s) < \epsilon, \]
and,
\[ f_{\pi} (s) = - \frac{\sigma_P (s)}{\sigma_A (s)} \tmop{defined} \tmop{for}
   - \pi \leqslant \Im (s) < - \pi + \epsilon \tmop{or} \pi - \epsilon < \Im
   (s) \leqslant \pi \]
The precise value of $\epsilon > 0$ is not important but we do want to choose
$\epsilon < \pi / 2$ to stay away from the zeros of $\sigma_A (s)$ and
$\sigma_P (s)$. \ The functions $\sigma_A (s)$ and $\sigma_P (s)$ are $2 \pi
i$ periodic in $s$ so we may regard $f_0 $ as a function on $\Sigma (
\mathcal{D})$ defined in a neighborhood of $\Im (x) = 0$ and $f_{\pi}$ as a
function on $\Sigma ( \mathcal{D})$ defined in a neighborhood of $\Im (x) =
\pm \pi$.

The change of sign in the definition of $f_{\pi}$ is important because of the
particular technique we use to produce a holomorphic factorization of the pair
$(f_0, f_{\pi})$ on the spectral curve. \ As defined $f_0 (s)$ tends to 1 as
$\Re (s) \rightarrow \pm \infty$ on the line $\Im (s) = 0$ and $f_{\pi} (s)$
tends to 1 as $\Re (s) \rightarrow \pm \infty$ on the line $\Im (s) = \pm
\pi$. \ In fact, it is useful to note that,
\begin{eqnarray}
  f_0 (s) = 1 + O (e^{- m L \tmop{ch} s}) \tmop{for} \Im (s) = 0 \tmop{and} s
  \rightarrow \pm \infty &  &  \nonumber\\
  f_{\pi} (s + i \pi) = 1 + O (e^{- m L \tmop{ch} s}) \tmop{for} \Im (s) = 0
  \tmop{and} s \rightarrow \pm \infty &  &  \label{fasymp}
\end{eqnarray}
As a consequence both $\log (f_0)$ and $\log (f_{\pi})$ will tend to 0 as $\Re
(s) \rightarrow \pm \infty$ and this will insure the convergence of the
integrals we use to define the additive splitting of the logarithms. \ In
fact, since neither $f_0$ or $f_{\pi}$ has either a zero or a pole in its
simply connected domain of definition they both possess holomorphic logarithms
which we can normalize to be 0 at $\infty$ in $s$. \ We henceforth denote
these choices by $\log f_0$ and $\log f_{\pi}$. \ It follows from
(\ref{fasymp}) that,
\begin{eqnarray}
  \log f_0 (s) = O (e^{- m L \tmop{ch} s}) \tmop{for} \Im (s) = 0 \tmop{and} s
  \rightarrow \pm \infty &  &  \nonumber\\
  \log f_{\pi} (s + i \pi) = O (e^{- m L \tmop{ch} s}) \tmop{for} \Im (s) = 0
  \tmop{and} s \rightarrow \pm \infty &  &  \label{logasymp}
\end{eqnarray}
We can use the Green function for the Cauchy--Riemann operator on the spectral
curve, $\Sigma ( \mathcal{D})$, to additively split this pair in the usual
fashion. \ Define the level sets for the imaginary part of $s$,
\[ Y_a =\{s | \Im (s) = a\} \]
positively oriented by the one form $d s$. \ For $0 < \Im (x) < \pi$ define,
\begin{equation}
  F_+ (z) = \frac{1}{2 \pi i} \int_{Y_0} \frac{e^s}{e^s - e^z} \log \bignone
  f_0 (s) d s - \frac{1}{2 \pi i} \int_{Y_{\pi}} \frac{e^s}{e^s - e^z}
  \bignone \log f_{\pi} (s) d s. \label{eq:7}
\end{equation}
For $- \pi < \Im (z) < 0$ define,
\begin{equation}
  F_- (z) = \frac{1}{2 \pi i} \int_{Y_{- \pi}} \frac{e^s}{e^s - e^z} \log
  f_{\pi} (s) d s - \frac{1}{2 \pi i} \int_{Y_0} \frac{e^s}{e^s - e^z} \log
  f_0 (s) d s. \bignone \bignone \label{eq:8}
\end{equation}
Because all the integrals involved converge absolutely, the functions $F_{\pm}
(z)$ are holomorphic in their domains of definition. \ In fact, since
\[ \frac{1}{2 \pi i} \frac{e^s}{e^s - e^z} \]
is the kernal of the Green function for the Cauchy--Riemann operator on the
spectral curve one can see that $F_+ (z)$ and $F_- (z)$ have boundary values
such that,
\[ F_+ (z) + F_- (z) = \log f_0 (z) \tmop{for} \Im (z) = 0 \]
and
\[ F_+ (z) + F_- (z) = \log f_{\pi} (z) \tmop{for} \Im (z) = \pm \pi, \]
where $Y_{\pi}$ is identified with $Y_{- \pi}$ in the second equality of
boundary values.

Once one knows this, it is clear that both $F_+ (z)$ and $F_- (z)$ extend
holomorphically to the strips $- \epsilon < \Im (z) < \pi + \epsilon$ \ and $-
\pi - \epsilon < \Im (z) < \epsilon$ respectively since the domain of
analyticity for $\log f_0$ and $\log f_{\pi}$ naturally enlarges the domain of
analyticity for $F_{\pm}$. \ It is convenient to regard the enlarged domains
for $F_{\pm}$ as subsets of the periodic strip $\Sigma ( \mathcal{D})
=\tmmathbf{C}/ 2 \pi i\tmmathbf{Z}$. \ Thus both $F_+ (z)$ and $F_- (z)$ are
holomorphic in a neighborhood of both $\tmop{Im} (z) = 0$, and $\Im (z) = \pm
\pi$ in $\Sigma ( \mathcal{D})$.

Rewrite the formulas (\ref{eq:7}) and (\ref{eq:8}) using $s$ to parametrize
$Y_0$ and $s \pm i \pi$ to parametrize $Y_{\pm \pi}$. \ One finds,
\begin{equation}
  F_+ (z) = \frac{1}{2 \pi i} \int_{- \infty}^{\infty} \frac{1}{\tmop{sh} (z -
  s)} \bignone \log \frac{\sigma_P (s)}{\sigma_A (s)} d s, \tmop{for} 0 < \Im
  (z) < \pi \label{eq:9}
\end{equation}
and
\begin{equation}
  F_- (z) = - \frac{1}{2 \pi i} \int_{- \infty}^{\infty} \frac{1}{\tmop{sh} (z
  - s)} \bignone \log \frac{\sigma_P (s)}{\sigma_A (s)} d s, \tmop{for} - \pi
  < \Im (z) < 0. \label{eq:10}
\end{equation}
Since the hyperbolic sine is $i \pi$ anti-periodic it follows from this
representation that,
\begin{equation}
  F_+ (z + i \pi) - F_- (z) = 0, \tmop{for} - \pi < \Im (z) < 0. \label{eq:11}
\end{equation}

Next we introduce the modifications needed to holomorphically factor $\sigma_P
/ \sigma_A$ with no sign change on $Y_{\pm \pi} .$ Define,
\begin{eqnarray}
  &  & \lambda_P (s) = e^{s / 2} \exp (F_+ (s)), \tmop{for} - \varepsilon <
  \Im (s) < \pi + \varepsilon . \nonumber\\
  &  & \lambda_A (s) = e^{s / 2} \exp (- F_- (s)), \tmop{for} - \pi -
  \varepsilon < \Im (s) < \varepsilon  \label{12}
\end{eqnarray}
We regard both $\lambda_P$ and $\lambda_A$ as holomorphic functions defined on
open subsets of $\Sigma ( \mathcal{D})$. \ Keep in mind, however, that because
$e^{s / 2}$ is $2 \pi i$ anti-periodic the representation for $\lambda_P (s)$
becomes,
\[ \lambda_P (s) = - e^{s / 2} \exp (F_+ (s)), \tmop{for} - \pi \leqslant \Im
   (s) < - \pi + \varepsilon, \]
with a similar modification in the formula for $\lambda_A (s)$ for $\pi -
\varepsilon < \Im (s) \leqslant \pi$. \

With this $2 \pi i$ anti-periodic adjustment one finds that,
\begin{eqnarray*}
  \frac{\lambda_P (s)}{\lambda_A (s)} = \frac{\sigma_P (s)}{\sigma_A (s)} &
  \tmop{for} s \tmop{in} a \tmop{neighborhood} \tmop{of} Y_0 \tmop{and}
  Y_{\pi} = Y_{- \pi} (\tmop{in} \Sigma ( \mathcal{D})), & 
\end{eqnarray*}
or more conveniently for us,
\begin{equation}
  \frac{\lambda_P (s)}{\sigma_P (s)} = \frac{\lambda_A (s)}{\sigma_A (s)}
  \tmop{for} s \tmop{near} Y_0 \tmop{and}^{} Y_{\pi} = Y_{- \pi} (\tmop{in}
  \Sigma ( \mathcal{D})) . \label{eq:13}
\end{equation}

\section{The Green function $\mathcal{G}_{P / A} (x, x')$}

Following Lisovyy {\cite{L05}} we write down a Green function for the Dirac
operator on $[- L, L] \times \tmmathbf{R}$ which has periodic boundary
conditions on $[- L, L] \times \tmmathbf{R}_+$ and anti-periodic boundary
conditions on $[- L, L] \times \tmmathbf{R}_-$, where $\tmmathbf{R}_+$ is the
set of positive real numbers and $\tmmathbf{R}_-$ is the set of negative real
numbers. Define,
\[ Y^a_b = Y_{b - a} - Y_{b + a}, \]
which is the ``counterclockwise'' oriented boundary of the strip $b - a
\leqslant \Im (z) \leqslant b + a$, centered at $Y_b$ and having width $2 a$.
\ Next we introduce partial definitions for the Green function $\mathcal{G}_{P
/ A} (x, x')$ of principal interest for us. \ For $\Im (x) > 0 > \Im (x')$
define,
\begin{equation}
  \mathcal{G}_{P / A} (x, x') = g_A (x - x') - \frac{i m}{32 \pi^2}
  \int_{Y_{\pi / 2}^a} d s \int_{Y_{- \pi / 2}^b} d t E (x, s) \frac{\lambda_P
  (s)}{\sigma_P (s)}  \frac{e^s - e^t}{e^s + e^t}  \bignone \bignone
  \frac{\lambda_A (t)}{\sigma_A (t)} E (x', t)^T . \label{eq:14}
\end{equation}
For $\Im (x) < 0 < \Im (x')$ define,
\begin{equation}
  \mathcal{G}_{P / A} (x, x') = g_P (x - x') - \frac{i m}{32 \pi^2} \int_{Y_{-
  \pi / 2}^a} d s \bignone \int_{Y_{\pi / 2}^b} d t \bignone E (x, s) \bignone
  \bignone \frac{\lambda_A (s)}{\sigma_A (s)} \frac{e^s - e^t}{e^s + e^t} 
  \frac{\lambda_P (t)}{\sigma_P (t)} E (x', t)^T . \label{eq:15}
\end{equation}

In these formulas we suppose that,
\[ 0 < b < a < \pi / 2. \]
In both integrals above the difference $\Im (s - t)$ is $\pi \pm (a - b)$ or
$\pi \pm (a + b)$ or $- \pi \pm (a - b)$ or $- \pi \pm (a + b)$. \ In no case
is $e^s + e^t = 0$ since this would require that $s$ and $t$ have imaginary
parts that differ by some odd multiple of $\pi$. \ Also, the restrictions on
$a$ and $b$ keep $s$ and $t$ away from the zeros of $\sigma_P$ and $\sigma_A$.
Thus the integrals \ do not have any local singularities. \ Next we check
convergence for large values of $s$ and $t$. \ It will suffice to illustrate
the estimate for the integrand in the $s$ integration in (\ref{eq:14}). \
Write $s = u + i v$, and $x = x_1 + i x_2$. Using the easily confirmed
observation that $F_{\pm} (s)$ tends to 0 as $|s| \rightarrow \infty$ we find
the asymptotics ($u \rightarrow \pm \infty$),
\[ | \frac{\lambda_P (s)}{\sigma_P (s)} E (x, s) | = O \left( \frac{\sqrt{e^u
   + e^{- u}}}{e^{m L \tmop{ch} (u)}} \exp (m x_1 \tmop{ch} (u) \cos (v) - m
   x_2 \tmop{ch} (u) \sin (v)) \right) . \]
As long as $x_2 \geqslant 0, \sin (v) \geqslant 0,$ and $|x_1 | < L$, it
follows that, for any $\varepsilon$ such that $0 < \varepsilon < m (L - x_1
\cos (v))$,
\[ | \frac{\lambda_P (s)}{\sigma_P (s)} E (x, s) | = O \left( \exp (-
   \varepsilon \tmop{ch} (u)) \right) . \]
Since $\sin (\pi / 2 \pm a) \geqslant 0$ for $0 < a \leqslant \pi / 2$, the
$s$ integral in (\ref{eq:15}) converges at $\infty$ provided only that $x_2
\geqslant 0$ and $|x_1 | < L$ (this is true even for the limiting case $a =
\pi / 2$). \ In a similar fashion one can see that for $x' = x_1' + i x_2'$
the $t$ integral in (\ref{eq:15}) converges at $\infty$ provided $x_2'
\leqslant 0$ and $|x_1' | < L$ (again even for $a = \pi / 2$).

This is useful since we understand (\ref{eq:14}) and (\ref{eq:15}) in two
ways. \ First by taking limits $a \rightarrow \pi / 2$ and $b \rightarrow \pi
/ 2$, and then by collapsing the integrals to the residues at the zeros of
$\sigma_P$ and $\sigma_A$.

First we take the limit $a \rightarrow \pi / 2$ of the right hand side of
(\ref{eq:14}) with $b$ fixed. \ The analyticity of the integrand in $s$ and
the asymptotics for large $s$ justify the replacement $Y_{\pi / 2}^a
\rightarrow Y_0 - Y_{\pi}$ in this limit. \ We cannot do the same with the
limit $b \rightarrow \pi / 2$ since the points $s \in Y_{\pi}$ and $t = s - i
\pi \in Y_0$ and $s \in Y_0$ and $t = s - i \pi \in Y_{- \pi}$ \ are zeros of
$e^s + e^t$. \ In the limit $b \rightarrow \pi / 2$it is possible to skirt
round $t = s - i \pi$ with a half circle of radius $\varepsilon$. \ The limit
$\varepsilon \rightarrow 0$ then falls into two pieces. \ The first is a
principal value integral and the second is a half residue. \ The principal
value contribution to the limiting value of the integral on the right hand
side of (\ref{eq:14}) is,
\[ - \lim_{\varepsilon \rightarrow 0}  \frac{i m}{32 \pi^2} \int_{\Gamma
   (\varepsilon)} d s d t E (x, s) \bignone \frac{\lambda_P (s)}{\sigma_P (s)}
   \frac{e^s - e^t}{e^s + e^t}  \bignone \bignone \frac{\lambda_A
   (t)}{\sigma_A (t)} E (x', t)^T, \]

where,
\[ \Gamma (\varepsilon) = (Y_0 - Y_{\pi}) \times (Y_{- \pi} - Y_0) \backslash
   \{(s, t) : |t - s + i \pi | < \varepsilon\} \]
The half residue contribution is,
\[ \frac{m}{16 \pi} \int_{Y_0 - Y_{\pi}} d s E (x, s) \frac{\lambda_P (s)
   \lambda_A (s - i \pi)}{\sigma_P (s) \sigma_A (s - i \pi)} \bignone E (x', s
   - i \pi)^T . \]
Consulting (\ref{eq:11}) we see that,
\[ \lambda_P (s) \lambda_A (s - i \pi) = - i e^s, \]
so that this residue contribution is,
\[ - \frac{i m}{16 \pi} \int_{Y_0 - Y_{\pi}} d s E (x, s) \bignone 
   \frac{e^s}{\sigma_P (s) \sigma_A (s - i \pi)} E (x', s - i \pi)^T . \]
This integral collapses to a sum over the residues at $s \in \Sigma_P' + i \pi
/ 2$ \ and $s \in \Sigma_A' + i \pi / 2$. One finds (recall that $s_{\pm} = s
\pm i \pi / 2$),
\[ \frac{1}{8 L} \sum_{s \in \Sigma_P'} \frac{e^s}{\tmop{ch} s} E (x, s_+) E
   (x', s_-)^T \bignone - \frac{1}{8 L} \sum_{s \in \Sigma_A'}
   \frac{e^s}{\tmop{ch} s} E (x, s_+) E (x', s_-)^T_{\bignone} \bignone, \]
or consulting (\ref{eq:5}),
\[ \frac{1}{2} g_P (x - x') - \frac{1}{2} g_A (x - x') . \]
Thus for $\Im (x) > 0 > \Im (x')$,
\begin{equation}
  \mathcal{G}_{P / A} (x, x') = g_{P + A} (x - x') - \lim_{\varepsilon
  \rightarrow 0}  \frac{i m}{32 \pi^2} \int_{\Gamma (\varepsilon)} d s d t E
  (x, s) \bignone \frac{\lambda_P (s)}{\sigma_P (s)}  \frac{e^s - e^t}{e^s +
  e^t}  \bignone \bignone \frac{\lambda_A (t)}{\sigma_A (t)} E (x', t)^T,
  \label{eq:16}
\end{equation}
where,
\[ g_{P + A} = (g_P + g_A) / 2. \]
A precisely analogous calculation shows that for $\Im (x) < 0 < \Im (x')$,
\[ \mathcal{G}_{P / A} (x, x') = g_{P + A} (x - x') - \lim_{\varepsilon
   \rightarrow 0}  \frac{i m}{32 \pi^2} \int_{\Gamma' (\varepsilon)} d s d t E
   (x, s) \bignone \frac{\lambda_A (s)}{\sigma_A (s)}  \frac{e^s - e^t}{e^s +
   e^t}  \frac{\lambda_P (t)}{\sigma_P (t)}  \bignone \bignone E (x', t)^T, \]
where $\Gamma' (\varepsilon) = (Y_{- \pi} - Y_0) \times (Y_0 - Y_{\pi})
\backslash \{(s, t) : |t - s - i \pi | < \varepsilon\}$. We transform the last
integral using (\ref{eq:13}); we replace $\lambda_P (t) / \sigma_P (t)$ by
$\lambda_A (t) / \sigma_A (t)$ and $\lambda_A (s) / \sigma_A (s)$ by
$\lambda_P (s) / \sigma_P (s)$ and then use $Y_{\pi} = Y_{- \pi}$ on $\Sigma (
\mathcal{D})$, to replace $\Gamma' (\varepsilon)$ with $\Gamma (\varepsilon)$.
\ One finds that for $\Im (x) < 0 < \Im (x')$,
\begin{equation}
  \mathcal{G}_{P / A} (x, x') = g_{P + A} (x - x') - \lim_{\varepsilon
  \rightarrow 0}  \frac{i m}{32 \pi^2} \int_{\Gamma (\varepsilon)} d s d t E
  (x, s) \bignone \frac{\lambda_P (s)}{\sigma_P (s)}  \frac{e^s - e^t}{e^s +
  e^t}  \bignone \bignone \frac{\lambda_A (t)}{\sigma_A (t)} E (x', t)^T .
  \label{eq:17}
\end{equation}
Note that in the formulas (\ref{eq:16}) and (\ref{eq:17}) the integrals on the
right hand sides now have the same shape. This has important consequences for
a polarization that we will associate with the kernel $\mathcal{G}_{P / A} (x,
x')$.

We turn to the other evaluation of $\mathcal{G}_{P / A} (x, x')$ that is
important for us. \ \ We can evaluate the integrals in (\ref{eq:14}) by
residues at the poles on $s \in Y_{\pi / 2}$ and $t \in Y_{- \pi / 2}$ not
forgeting the poles in $(e^s + e^t)^{- 1}$ that arise for $s = t + i \pi$
(this pole gives rise to $- g_A (x - x')$). \ Since we want to maintain $a >
b$, the $t$integral residue calculation should be done first. \ \ For $\Im (x)
> 0 > \Im (x')$ we find,
\[ \mathcal{G}_{P / A} (x, x') = \frac{i}{8 m L^2} \sum_{s \in \Sigma^+_P}
   \bignone \sum_{t \in \Sigma^-_A} E (x, s) \frac{\lambda_P (s)}{\tmop{sh} s}
   \frac{e^s - e^t}{e^s + e^t}  \frac{\lambda_A (t)}{\tmop{sh} t} e^{- m L
   (\tmop{ch} s + \tmop{ch} t)} E (x', t)^T . \]
where,
\[ \Sigma_{P, A}^{\pm} = \Sigma'_{P, A} \pm i \pi / 2. \]
Recalling that $s_{\pm} = s \pm i \pi / 2$, we can turn this into a sum over
real arguments, which makes it easier to see the convergence of the sum. For
$\Im (x) > 0 > \Im (x')$,
\begin{equation}
  \mathcal{G}_{P / A} (x, x') = \frac{i}{8 m L^2} \sum_{s \in \Sigma'_P}
  \sum_{t \in \Sigma'_A} E (x, s_+) \frac{\lambda_P (s_+)}{\tmop{ch} s}
  \frac{e^s + e^t}{e^s - e^t} \frac{\lambda_A (t_-)}{\tmop{ch} t} e^{i m L
  (\tmop{sh} t - \tmop{sh} s)} E (x', t_-)^T \label{eq:18}
\end{equation}

An analogous residue calculation for $\Im (x) < 0 < \Im (x')$ shows that,
\[ \mathcal{G}_{P / A} (x, x') = \frac{i}{8 m L^2} \sum_{s \in \Sigma^-_A}
   \bignone \sum_{t \in \Sigma^+_P} E (x, s) \frac{\lambda_P (s)}{\tmop{sh} s}
   \frac{e^s - e^t}{e^s + e^t}  \frac{\lambda_A (t)}{\tmop{sh} t} e^{- m L
   (\tmop{ch} s + \tmop{ch} t)} E (x', t)^T \]
And again this transforms into a sum on real arguments. For $\Im (x) < 0 < \Im
(x')$,
\begin{equation}
  \mathcal{G}_{P / A} (x, x') = \frac{i}{8 m L^2} \sum_{s \in \Sigma'_A}
  \bignone \sum_{t \in \Sigma'_P} E (x, s_-) \frac{\lambda_A (s_-)}{\tmop{ch}
  s} \frac{e^s + e^t}{e^s - e^t}  \frac{\lambda_P (t_+)}{\tmop{ch} t} e^{i m L
  (\tmop{sh} s - \tmop{sh} t)} E (x', t_+)^T \label{eq:19}
\end{equation}
The reader can now check that the representations (\ref{eq:16}) and
(\ref{eq:18}) for $\mathcal{G}_{P / A} (x, x')$ extend continuously from
the domain $\Im (x) > 0 > \Im (x')$ to $\Im (x) > 0 \geqslant \Im (x')$ and
the representations (\ref{eq:17}) and (\ref{eq:19}) for
$\mathcal{G}_{P / A} (x, x')$ extend continously from the domain
$\Im (x) < 0 < \Im (x')$ to the domain $\Im (x) < 0 \leqslant \Im (x')$. \ In
each case, we further require that $|x_1 | < L$ and $|x_1' | < L$ to guarentee
that the simple estimates we gave above assure pointwise convergence in the
integral representations (the sums don't have a problem).

Associated with the Green function $\mathcal{G}_{P / A}$ there is a splitting
of $L^2 [- L, L]$ which is important for us. \ We will first discuss this in a
naive setting and then introduce the estimates needed to make it work
mathematically. \ We hope this will make it easier for the reader to see the
reason for the definitions (\ref{eq:14}) and (\ref{eq:15}). \ For $x, x' \in
[- L, L]$ define,
\begin{equation}
  \mathcal{P}_{P / A}^+ f (x) = \lim_{\varepsilon \downarrow 0} \int_{- L}^L
  \mathcal{G}_{P / A} (x + i \varepsilon, x') f (x') d x' \bignone
  \text{\label{eq:20}}
\end{equation}
and
\begin{equation}
  \mathcal{P}_{P / A}^- f (x) = - \lim_{\varepsilon \downarrow 0} \int_{- L}^L
  \mathcal{G}_{P / A} (x - i \varepsilon, x') \bignone f (x') d x'
  \label{eq:21}
\end{equation}
Note that (\ref{eq:18}) and (\ref{eq:19}) suggest that,
\begin{equation}
  \mathcal{P}_{P / A}^+ f \in W_P^+, \tmop{and} \mathcal{P}_{P / A}^- f \in
  W_A^- . \label{split}
\end{equation}
On the other hand (\ref{eq:16}) and (\ref{eq:17}) suggest that for $\Im (x) =
\Im (x') = 0,$
\begin{equation}
  \mathcal{G}_{P / A} (x + i 0, x') - \mathcal{G}_{P / A} (x - i 0, x') = g_{P
  + A} (x - x' + i 0) - g_{P + A} (x - x' - i 0) = \delta (x - x'),
  \label{delta}
\end{equation}
since,
\[ g_P (x - x' + i 0) - g_P (x - x' - i 0) = g_P (x - x' + i 0) - g_P (x - x'
   - i 0) = \delta (x - x') . \]
Thus we expect that,
\begin{equation}
  \mathcal{P}_{P / A}^+ f + \mathcal{P}_{P / A}^- f = f. \label{split1}
\end{equation}
We will prove later that,
\[ \mathcal{P}_{P / A^{}}^{\pm} \mathcal{P}_{P / A}^{\mp} = 0. \]
Combined with (\ref{split1}) this implies that $\mathcal{P}_{P / A^{}}^{\pm}$
is a projection. These projections will allow us to explicitly invert the map,
\begin{equation}
  Q_P^- : W_A^- \rightarrow W_P^- . \label{eq:22}
\end{equation}
Note that if,
\[ \left(\begin{array}{cc}
     A & B\\
     C & D
   \end{array}\right), \]
is the matrix of the identity map on $W$ from the splitting $W_A^+ \oplus
W_A^-$ to $W_P^+ \oplus W_P^-$ then (\ref{eq:22}) is just the map $D$.

To see how to invert $D$ suppose that $f \in W_A^-$ and $Q_P^- f = g \in
W_P^-$. \ Then $f = g_P^+ + g$ where, $g_P^+ \in W_P^+$, and so it follows
from (\ref{split}), (\ref{split1}) and the fact that $\mathcal{P}^-_{P / A}$
is a projection that is 0 on $W_P^+$ that,
\begin{equation}
  f = \mathcal{P}^-_{P / A} f = \mathcal{P}_{P / A}^- (g_P^+ + g) =
  \mathcal{P}_{P / A}^- g = D^{- 1} g \label{eq:23}
\end{equation}
Thus (\ref{eq:23}) gives an explicit formula for the inversion of
(\ref{eq:22}). \

There are a number of matters that we glossed over in these ``pointwise''
calculations. \ The first matter we take up is a proof that $\mathcal{P}_{P /
A}^{\pm}$ is bounded in $L^2$. \ It is enough to illustrate this for
$\mathcal{P}_{P / A}^+$ using the formula (\ref{eq:18}). \ Suppose that $f$ is
a smooth compactly supported function on $[- L, L]$. \ Define,
\[ F (x) = \frac{i}{8 m L^2} \int_{- L}^L d x' \bignone \sum_{s \in \Sigma'_P}
   \bignone \sum_{t \in \Sigma'_A} E (x, s_+) \frac{\lambda_P (s_+)}{\tmop{ch}
   s} \frac{e^s + e^t}{e^s - e^t}  \frac{\lambda_A (t_-)}{\tmop{ch} t} E (x',
   t_-)^T f (x') . \]
Because we are interested in $L^2$ bounds we can and do ignore the unitary
factor $e^{i m L (\tmop{sh} t - \tmop{sh} s)}$ in (\ref{eq:18}). Using the
formula (\ref{spec}) for the spectral tranform and the formulas (\ref{12}) for
$\lambda_P$ and $\lambda_A$ we find,
\[ F (x) = \frac{i}{4 m L} \bignone \sum_{s \in \Sigma'_P} \bignone \sum_{t
   \in \Sigma'_A} E (x, s_+) e^{s / 2} \frac{e^{F_+ (s_+)}}{\tmop{ch} s}
   \frac{e^s + e^t}{e^s - e^t}  \frac{e^{- F_- (t_-)}}{\tmop{ch} t} 
   \mathcal{S}_P^+ f (t) . \]

Introduce,
\begin{equation}
  g (s) = \frac{i}{2 m L} \bignone \sum_{t \in \Sigma'_A} e^{F_+ (s_+)}
  \frac{e^s + e^t}{e^s - e^t}  \frac{e^{- F_- (t_-)}}{\tmop{ch} t}
  \mathcal{S}_P^+ f (t) . \label{specsim}
\end{equation}
Then the formula (\ref{inspec1}) implies that,
\[ F (x) = \mathcal{R}_P^+ g (s) . \]
Because $\mathcal{R}_P^+$ is an isometry it will suffice to obtain a suitable
estimate for,
\[ \sum_{s \in \Sigma_P'} \frac{|g (s) |^2}{2 \tmop{ch} s} \bignone . \]
However, since $\frac{e^s + e^t}{e^s - e^t}$, and $e^{F_+ (s_+) - F_- (t_-)}$
are uniformly bounded in $s \tmop{and} t$ for $s \in \Sigma_P'$ and $t \in
\Sigma_A' \bignone$, and $(\tmop{ch} t)^{- 1}$ is summable for $t \in
\Sigma_P'$ we can use the Cauchy-Schwartz inequality in (\ref{specsim}) to see
that,
\[ \sum_{s \in \Sigma_P'} \frac{|g (s) |^2}{2 \tmop{ch} s} \leqslant C \sum_{s
   \in \Sigma_P'} \frac{1}{\tmop{ch} s} \bignone \sum_{t \in \Sigma_P'}
   \frac{| \mathcal{S}_P^+ f (t) |^2}{2 \tmop{ch} t} \bignone . \]

This finishes the proof that $\mathcal{P}_{P / A}^+$ is bounded in $L^2$ on
the smooth functions of compact support in $[- L, L] .$ It then extends by
density and continuity to a continuous map on all of $L^2$.

Next we show that $\mathcal{P}_{P / A}^+ f \in W_P^+$. Truncating the $s$ sum
to a finite range in the contribution that the double sum in (\ref{eq:18})
makes to the calculation of (\ref{eq:20}) we see that the result is clearly in
$W_P^+$. \ Since the $s$ sum converges in $L^2$ we see that $\mathcal{P}_{P /
A}^+ f \in W_P^+ .$ \ The same argument shows that $\mathcal{P}_{P / A}^- f
\in W_A^-$.

One other unresolved issue we wish to consider is the ``pointwise''
cancellation that produced (\ref{delta}). \ We do not have estimates for the
terms that cancelled at the endpoints $x = \pm L$ and $x' = \pm L$. \ However,
the $L^2$ continuity of the maps $\mathcal{P}_{P / A}^{\pm}$ makes it natural
to define them as $L^2 $limits defined first for functions, $f,$ of compact
support on $[- L, L]$. \ If $f$ is such a function then the contribution made
by the terms $g_{A + P}$ in (\ref{eq:16}) and (\ref{eq:17}) to $\mathcal{P}_{P
/ A}^+ f + \mathcal{P}_{P / A}^- f$ is,
\[ \frac{1}{2} \left( Q_P^+ + Q_A^+ \right) f + \frac{1}{2} \left( Q_P^- +
   Q_A^- \right) f = f. \]
Applied to a function of compact support in $[- L, L]$ the issue of
convergence for the $x'$ integration for the residual kernels in (\ref{eq:16})
and (\ref{eq:17}) in the calculation of $\mathcal{P}_{P / A}^{\pm} f$does not
encounter any difficulties at $x' = \pm L$. The resulting functions clearly
cancel except possibly at $x = \pm L$. \ However, since the kernels define
bounded operators on $L^2$ the cancelation is good in $L^2$ and it follows
that,
\[ \mathcal{P}_{P / A}^+ f + \mathcal{P}_{P / A}^- f = f \tmop{in} L^2 . \]
Finally we take up the proof that $\mathcal{P}_{P / A}^-$ maps $W_P^+$ to 0. \
Suppose that,
\[ f (x) = \mathcal{R}_P^+ F (x), \]
where $F (s)$ is non zero for only finitely many $s \in \Sigma_P'$. \ Such $f$
are dense in $W_P^+$ so it will suffice to prove that $\mathcal{P}_{P / A}^- f
= 0.$ \ Use (\ref{eq:19}) in the definition (\ref{eq:21}). \ \ One encounters
the integral,
\[ \int_{- L}^L E (x', t_+)^T f (x') d x' \bignone . \]
This can be rewritten as the integral of a one form,
\[ \int_{- L}^L E_1 (x', t_+) f_1 (x') d x' + E_2 (x', t_+) f_2 (x') d
   \bar{x}' . \bignone \]
It is not hard to check that as a consequence of the fact that both $x'
\rightarrow E (x', t_+)$ and $x' \rightarrow f (x')$ are well behaved
solutions to the Dirac equation in the upper half plane, the integrand in this
last integral is a closed one form. \ This means that the contour in the
integral can be shifted from $\Im (x') = 0$ to $\Im (x') = R > 0$ for any $R$.
\ However, both solutions tend to 0 as $R \rightarrow + \infty$ and it follows
that the integral itself must vanish. \ This shows that $\mathcal{P}_{P / A}^-
f = 0$ and finishes the proof. \ The only change needed in the proof that
$\mathcal{P}_{P / A}^+$ maps $W_A^-$ to 0 is the observation that the product
of two anti-periodic functions is periodic. \ This allows the contour
deformation argument to proceed.

Next we turn to an an extension of the definition of $\mathcal{G}_{P / A} (x,
x')$ to the remaining possible values for the argument $(x, x')$. \ First we
consider the extension of $\mathcal{G}_{P / A} (x, x')$ to $\Im (x) > \Im (x')
\geqslant 0.$ \ Start with (\ref{eq:18}). In that equation replace the $t$ sum
by a contour integral on the boundary of tubular neighborhood of $\Im (t) = -
\pi / 2$ using the kernel $\sigma_P (t)^{- 1}$. \ Note that the contour
integral that it is natural to introduce contains an ``extra'' pole at the
zero of $e^s + e^t$. We compensate by adding $g_P (x - x')$ to the contour
integral to obtain a precise representation of the original sum. \ Suppose
that $\Im (x) > 0 \tmop{and} \Im (x') = 0.$ Extend the contours in the $t$
integral to the horizontal lines $\Im (t) = 0, - \pi$. \ Replace the ratio
$\lambda_A (t) / \sigma_A (t)$ in the integrand by $\lambda_P (t) / \sigma_P
(t)$ using equation (\ref{eq:13}). \ Then collapse the resulting integral to
the sum of the residues on $\Im (t) = \pi / 2.$ Keeping in mind that the
orientation of the contour flips going from $\Im (t) = - \pi / 2$ to $\Im (t)
= \pi / 2$ one finds,
\begin{equation}
  \mathcal{G}_{P / A} (x, x') = g_P (x - x') - \frac{i}{8 m L^2} \sum_{s \in
  \Sigma^+_P} \bignone \sum_{t \in \Sigma^+_P} E (x, s) \frac{\lambda_P
  (s)}{\tmop{sh} s} \frac{e^s - e^t}{e^s + e^t}  \frac{\lambda_P
  (t)}{\tmop{sh} t} e^{- m L (\tmop{ch} s + \tmop{ch} t)} E (x', t)^T .
  \label{gext1}
\end{equation}
In this form the second term on the right hand side has a continuous extension
to $\Im (x') \geqslant 0.$ \ A little thought shows that the right hand side
of (\ref{gext1}) represents the Green function of interest when both $x$ and
$x'$ are in the upper half plane. \ An almost identical argument produces the
following formula for $\mathcal{G}_{P / A} (x, x')$ when both $x$ and $x'$ are
in the lower half plane,
\begin{equation}
  \mathcal{G}_{P / A} (x, x') = g_A (x - x') - \frac{i}{8 m L^2} \sum_{s \in
  \Sigma^-_A} \bignone \sum_{t \in \Sigma^-_A} E (x, s) \frac{\lambda_A
  (s)}{\tmop{sh} s} \frac{e^s - e^t}{e^s + e^t}  \frac{\lambda_A
  (t)}{\tmop{sh} t} e^{- m L (\tmop{ch} s + \tmop{ch} t)} E (x', t)^T .
  \label{gext2}
\end{equation}
We finish this section by using the projections $\mathcal{P}_{P / A}^{\pm}$ to
give formulas for $B D^{- 1}$, and $D^{- 1} C$. \ Observe first that since,
\[ \left(\begin{array}{cc}
     A & B\\
     C & D
   \end{array}\right), \]
is the matrix of the identity, every element $f \in W_A^-$ can be written,
\[ f = D f + B f, \]
or since $D$ is invertible,
\[ W_A^- \ni f = g + B D^{- 1} g, \tmop{where} g \in W_P^- \tmop{and} B D^{-
   1} g \in W_P^+ . \]
This is described by saying that $W_A^-$ is the graph of $B D^{- 1} : W_P^-
\rightarrow W_P^+$ over $W_P^- .$ \ Now apply $\mathcal{P}_{P / A}^+$ to both
sides of this last equation and use the fact that $W_A^-$ is in the kernel of
this map and that the elements in $W_P^+$ are fixed by this map. \ One finds,
\[ 0 = \mathcal{P}_{P / A}^+ g + B D^{- 1} g \tmop{for} g \in W_P^- . \]
Thus,
\begin{equation}
  B D^{- 1} g = - \mathcal{P}_{P / A}^+ g \tmop{for} g \in W_P^-
  \label{bdinverse}
\end{equation}
Note that (\ref{gext1}) is a particularly effective representation for
$\mathcal{G}_{P / A}$ in the evaluation of this representation for $B D^{-
1}$. \ In particular the $g_P$ term in (\ref{gext1}) makes no contribution
since $Q_P^+ g = 0$ for $g \in W_P^-$.

Next we consider a representation for $D^{- 1} C$. \ Observe first that since
the identity map is an orthogonal map we have,
\begin{equation}
  \left(\begin{array}{cc}
    A & B\\
    C & D
  \end{array}\right) \left(\begin{array}{cc}
    D^{\tau} & B^{\tau}\\
    C^{\tau} & A^{\tau}
  \end{array}\right) = \left(\begin{array}{cc}
    1 & 0\\
    0 & 1
  \end{array}\right), \label{idorth}
\end{equation}
so that,
\[ \left(\begin{array}{cc}
     D^{\tau} & B^{\tau}\\
     C^{\tau} & A^{\tau}
   \end{array}\right), \]
is the matrix of the identity from $W = W_P^+ \oplus W_P^-$ to $W = W_A^+
\oplus W_A^-$. \ Thus for $f \in W_P^+$ we have,
\[ W_P^+ \ni f = D^{\tau} f + C^{\tau} f, \]
or introducing $g = D^{\tau} f \in W_A^+$ we have,
\[ W_P^+ \ni f = g + C^{\tau} D^{- \tau} g \tmop{for} g \in W_A^+ . \]
Using the identity $C D^{\tau} + D C^{\tau} = 0$ that follows from
(\ref{idorth}) this becomes the graph representation,
\[ W_P^+ \ni f = g - D^{- 1} C g \tmop{for} g \in W_A^+ . \]
Applying $\mathcal{P}_{P / A}^-$ to both sides and noting that $W_P^+$ is in
the null space and $W_A^-$ is in the range of this projection we find,
\begin{equation}
  D^{- 1} C g = \mathcal{P}_{P / A}^- g \tmop{for} g \in W_A^+
  \label{dinversec}
\end{equation}

\section{Matrix for $D^{- 1}, B D^{- 1}, \tmop{and} D^{- 1} C$ in the spectral
variables}

In this section we translate the representations (\ref{eq:23}),
(\ref{bdinverse}) and (\ref{dinversec}) into the spectral variables
(\ref{spec}) and (\ref{inspec}). \ Then we make some adjustments in these
representations that we held off because incorporating them earlier on would
have forced us to keep track of branches for the square root of $\tmop{sh} t$,
\ or irritating factors of $\exp (i \pi / 4)$ in the calculations of the
preceeding sections.

We start with the calculations for $D^{- 1}$. \ Substitute (\ref{eq:19}) in
(\ref{eq:23}) and make obvious use of (\ref{spec}) and (\ref{inspec}). \ One
finds the matrix representation of $D^{- 1}$ is,
\begin{equation}
  D^{- 1} \sim e^{- m L \tmop{ch} s} \frac{e^{- s / 2} \lambda_A (s)}{2 m L} 
  \frac{e^s - e^t}{e^s + e^t}  \frac{e^{- s / 2} \lambda_P (t)}{\tmop{sh} t}
  e^{- m L \tmop{ch} t} \tmop{for} s \in \Sigma_A^-, \tmop{and} t \in
  \Sigma_P^+ \label{mdinv1}
\end{equation}
This should be understood in the following sense. \ Replace $s$ by $s_-$ with
the new value of $s$ chosen in $\Sigma_A'$. Replace $t$ with $t_+$ with the
new value of $t$ chosen in $\Sigma_P'$. The matrix of $D^{- 1}$ in
(\ref{mdinv1}) is implicitly indexed by $\Sigma_A' \times \Sigma_P'$.

Substituting (\ref{gext1}) in (\ref{bdinverse}) one finds (using the
convention just explained) that the matrix of $B D^{- 1}$ is,
\[  B D^{- 1} \sim - i e^{- m L \tmop{ch} s} \frac{e^{- s / 2} \lambda_P
   (s)}{2 m L}  \frac{e^s - e^t}{e^s + e^t}  \frac{e^{- s / 2} \lambda_P
   (t)}{\tmop{sh} t} e^{- m L \tmop{ch} t} \tmop{for} s \in \Sigma_P^+,
   \tmop{and} t \in \Sigma_P^+ . \]
Substitute (\ref{gext2}) in (\ref{dinversec}) and one finds for the matrix of
$D^{- 1} C,$
\[ D^{- 1} C \sim i e^{- m L \tmop{ch} s} \frac{e^{- s / 2} \lambda_A (s)}{2 m
   L}  \frac{e^s - e^t}{e^s + e^t}  \frac{e^{- s / 2} \lambda_A (t)}{\tmop{sh}
   t} e^{- m L \tmop{ch} t} \tmop{for} s \in \Sigma_A^-, \tmop{and} t \in
   \Sigma_A^- . \]
Next we get rid of the asymmetry in these results that comes from using the
$\ell^2$ space on the spectral side with weight $(2 \tmop{ch} t)^{- 1}$. \ One
can pass to the matrix representation in the unweighted $\ell^2$ space by
multiplying on the left by $(2 \tmop{ch} s)^{- \frac{1}{2}}$ and on the right
by $(2 \tmop{ch} t)^{\frac{1}{2}}$ where $(s, t) \in \Sigma_{A, P}' \times
\Sigma_{A, P}'$ in whichever combination is appropriate. \ Do this and then
adjust the spectral transform further by multiplying by $e^{i \pi / 4}$ in the
$W_{A, P}^-$ subspaces and by $e^{- i \pi / 4}$ in the $W_{A, P}^+$ subspaces.
One finds,

\begin{theorem}
  \label{matrixelements}Suppose that
  \[ \left(\begin{array}{cc}
       A & B\\
       C & D
     \end{array}\right) \]
  is the matrix of the identity map on $W$ from the $W_A^+ \oplus W_A^-$
  splitting to the $W_P^+ \oplus W_P^-$ splitting. \ Then relative to the
  {\tmem{modified}} spectral--inverse spectral transforms,
  \[ \begin{array}{ll}
       \mathcal{S}_{A, P}^+ f (s) = \frac{e^{- i \pi / 4}}{2 L} \int_{- L}^L
       \frac{e^{s / 2}}{\sqrt{2 \tmop{ch} s}} E (x, s_-)^T f (x) d x \bignone
       & \tmop{for} s \in \Sigma_{A, P}',\\
       \mathcal{S}_{A, P}^- f (s) = \frac{e^{i \pi / 4}}{2 L} \int_{- L}^L
       \frac{e^{s / 2}}{\sqrt{2 \tmop{ch} s}} E (x, s_+)^T f (x) d x \bignone
       & \tmop{for} s \in \Sigma_{A, P}' .
     \end{array} \]
  and
  \[ \begin{array}{l}
       f (x) = \sum_{s \in \Sigma_P'} \frac{e^{s / 2}}{\sqrt{2 \tmop{ch} s}}
       \bignone \left( e^{i \pi / 4} \mathcal{S}_P^+ f (s) E (x, s_+) + e^{- i
       \pi / 4} \mathcal{S}_P^- f (s) E (x, s_-) \right)\\
       f (x) = \sum_{s \in \Sigma_A'} \frac{e^{s / 2}}{\sqrt{2 \tmop{ch} s}}
       \bignone \left( e^{i \pi / 4} \mathcal{S}_A^+ f (s) E (x, s_+) + e^{- i
       \pi / 4} \mathcal{S}_A^- f (s) E (x, s_-) \right)
     \end{array} \]
  The matrix elements of $D^{- 1}$, $B D^{- 1}$ and $D^{- 1} C$ are given by,
\end{theorem}
\[ D^{- 1} \sim \frac{1}{2 m L} e^{- m L \tmop{ch} s_-} \frac{e^{- s_- / 2}
   \lambda_A (s_-)}{\sqrt{\tmop{ch} s}}  \frac{e^{s_-} - e^{t_+}}{e^{s_-} +
   e^{t_+}}  \frac{e^{- t_+ / 2} \lambda_P (t_+)}{\sqrt{\tmop{ch} t}} e^{- m L
   \tmop{ch} t_+} \tmop{for} (s, t) \in \Sigma_A' \times \Sigma_P', \]
\[ B D^{- 1} \sim - \frac{1}{2 m L} e^{- m L \tmop{ch} s_+} \frac{e^{- s_+ /
   2} \lambda_P (s_+)}{\sqrt{\tmop{ch} s}}  \frac{e^{s_+} - e^{t_+}}{e^{s_+} +
   e^{t_+}}  \frac{e^{- t_+ / 2} \lambda_P (t_+)}{\sqrt{\tmop{ch} t}} e^{- m L
   \tmop{ch} t_+} \tmop{for} (s, t) \in \Sigma_P' \times \Sigma_P', \]
\[ D^{- 1} C \sim - \frac{1}{2 m L} e^{- m L \tmop{ch} s_-} \frac{e^{- s_- /
   2} \lambda_A (s_-)}{\sqrt{\tmop{ch} s}}  \frac{e^{s_-} - e^{t_-}}{e^{s_-} +
   e^{t_-}}  \frac{e^{- t_- / 2} \lambda_A (t_-)}{\sqrt{\tmop{ch} t}} e^{- m L
   \tmop{ch} t_-} \tmop{for} (s, t) \in \Sigma_A' \times \Sigma_A' . \]
We did not bother to introduce new notation for the modified transforms that
appear in this theorem since these transforms (modified or not) will make no
further appearance. \ We note that the spectral transforms introduced in
theorem \ref{matrixelements} are equivalent to the introduction of coordinates
for the bases (in the {\tmem{original}} spectral transform),
\[ e_{A, P}^{\pm} (s) = t \rightarrow \frac{e^{\pm i \pi / 4}}{\sqrt{2
   \tmop{ch} s}} \left(\begin{array}{c}
     e^{s / 2}\\
     \mp i e^{- s / 2}
   \end{array}\right) \delta (t, s) \tmop{for} t, s \in \Sigma_{A, P}' . \]
We follow the developments that lead up to theorem \ref{th:2}. \ Suppose that
$s_j' \in \Sigma_P$ for $j = 1, \ldots, m.$ Define
\[ e_P^+ (\tmmathbf{s}') = e_P^+ (s_1') \wedge \cdots \wedge e_P^+ (s_m'),
   \tmop{where} \tmmathbf{s}= (s_1', \ldots, s_m') . \]
For $s_j \in \Sigma_A \tmop{for} j = 1, \ldots, n$ define,
\[ e_A^+ (\tmmathbf{s}) = e_A^+ (s_1) \wedge \cdots \wedge e_A^+ (s_n) . \]
We apply theorem \ref{th:2} to the calculation of the multiparticle matrix
elements for the spin operator.
\begin{equation}
  \frac{\langle e_P^+ (\tmmathbf{s}'), \sigma e_A^+ (\tmmathbf{s})
  \rangle}{\langle 0_P, \sigma 0_A \rangle} = \tmop{Pf}
  \left(\begin{array}{cc}
    R_{\tmmathbf{s}' \times \tmmathbf{s}'} & R_{\tmmathbf{s}' \times
    \tmmathbf{s}}\\
    R_{\tmmathbf{s} \times \tmmathbf{s}'} & R_{\tmmathbf{s} \times
    \tmmathbf{s}}
  \end{array}\right), \label{mpmatrix}
\end{equation}
where $\tmop{Pf}$ is the Pfaffian and the skew symmetric matrix R has matrix
elements,
\[ \left( R_{\tmmathbf{s}' \times \tmmathbf{s}'} \right)_{i, j} =
   \frac{\langle e_P^+ (s_i') \wedge e_P^+ (s_j'), \sigma 0_A \rangle}{\langle
   0_P, \sigma 0_A \rangle} = B D^{- 1}_{s_i', s_j'} \]

\[ \left( R_{\tmmathbf{s}' \times \tmmathbf{s}} \right)_{i, j} = - \left(
   R_{\tmmathbf{s} \times \tmmathbf{s}'} \right)_{j, i} = \frac{\langle e_P^+
   (s_i'), \sigma e_A^+ (s_j) \rangle}{\langle 0_P, \sigma 0_A \rangle} = D^{-
   \tau}_{s_i', s_j} \]
and
\[ \left( R_{\tmmathbf{s} \times \tmmathbf{s}} \right)_{i, j} = \frac{\langle
   0_P, \sigma e_A^+ (s_i) \wedge e_A^+ (s_j)}{\langle 0_P, \sigma 0_A
   \rangle} = D^{- 1} C_{s_i, s_j} \]
Now theorm \ref{th:2} does not apply because a unitary intertwining map,
$\sigma$, does not exist as a map from the Fock representation on $\tmop{Alt}
(W_A^+)$ to the Fock representation on $\tmop{Alt} (W_P^+)$. \ However, \ all
the difficulty with the map can be handled by composing it with positive
powers of the transfer matrix on both sides. \ The factor \ $\langle 0_P,
\sigma 0_A \rangle$ remains undefined, however. \ There are more and less
natural ways to fix this constant but the discussion makes more sense if one
understands how to make a choice for the wave function renormalization in the
control of the scaling limit. \ Since we do not want to take up this matter
here we will set $\langle 0_P, \sigma 0_A \rangle = 1$ (which is one sensible
way in which to scale the spin operator) and proceed with the evaluation of
(\ref{mpmatrix}) using the formulas for the matrix elements in theorem
\ref{matrixelements}. We begin by defining two $(m + n) \times (m + n)$
matrices. \ Let $\Lambda$ denote the $(m + n) \times (m + n)$ diagonal matrix
with non-zero entries,
\[ \Lambda_{j, j} = \frac{e^{- m L \tmop{ch} s'_{j +}}}{\sqrt{2 m L \tmop{ch}
   s'_j}} e^{- s'_{j +} / 2} \lambda_P (s'_{j +}) \tmop{for} j = 1, \ldots, m,
\]
and
\[ \Lambda_{j + m, j + m} = \frac{e^{- m L \tmop{ch} s_{j -}}}{\sqrt{2 m L
   \tmop{ch} s_j}} e^{- s_{j -} / 2} \lambda_A (s_{j -}) \tmop{for} j = 1,
   \ldots, n. \]
Next introduce,
\[ t_j = s'_{j -} = s'_j - i \pi / 2, \tmop{for} j = 1, \ldots, m, \]
and
\[ t_{j + m} = s_{j +} = s_j + i \pi / 2, \tmop{for} j = 1, \ldots, n. \]
Define and $(m + n) \times (m + n)$ matrix, $T$, by,
\[ T_{i, j} = \frac{e^{t_i} - e^{t_j}}{e^{t_i} + e^{t_j}} \tmop{for} i, j = 1,
   \ldots, m + n. \]
Then using the results of theorem \ref{matrixelements} we see that,
\[ \left(\begin{array}{cc}
     R_{\tmmathbf{s}' \times \tmmathbf{s}'} & R_{\tmmathbf{s}' \times
     \tmmathbf{s}}\\
     R_{\tmmathbf{s} \times \tmmathbf{s}'} & R_{\tmmathbf{s} \times
     \tmmathbf{s}}
   \end{array}\right) = \Lambda T \Lambda = \Lambda T \Lambda^{\tau}, \]
where the transpose $\Lambda^{\tau}$ of the diagonal matrix $\Lambda$ is, of
course equal to $\Lambda$. \ Thus we have,
\[ \tmop{Pf} \left(\begin{array}{cc}
     R_{\tmmathbf{s}' \times \tmmathbf{s}'} & R_{\tmmathbf{s}' \times
     \tmmathbf{s}}\\
     R_{\tmmathbf{s} \times \tmmathbf{s}'} & R_{\tmmathbf{s} \times
     \tmmathbf{s}}
   \end{array}\right) = \tmop{Pf} (\Lambda T \Lambda^{\tau}) = \det (\Lambda)
   \tmop{Pf} (T) . \]
See {\cite{JP06}}. \ The determinant of $\Lambda$ is just the product of the
entries on the diagonal. To finish this calculation we observe that the
Pfaffian of $T$ has a product form,
\begin{equation}
  \tmop{Pf} (T) = \prod^{m + n}_{i < j} \bignone \frac{e^{t_i} -
  e^{t_j}}{e^{t_i} + e^{t_j}} . \label{product}
\end{equation}
This formula is doubtless well known but can be confirmed in the following
manner. \ Let $Z$ be an $N \times N$ matrix with entries,
\[ Z_{i, j} = \frac{z_i - z_j}{z_i + z_j} . \]
Then $Z$ is a skew-symmetric matrix and,
\begin{equation}
  \tmop{Pf} (Z) = \prod^N_{i < j} \frac{z_i - z_j}{z_i + z_j} \bignone .
  \label{product1} \text{}
\end{equation}
Expand the left hand side by ``minors'' of the first row (there is a suitable
expansion for Pfaffians {\cite{JP06}}). Use this to determine the residues at
the poles ($- z_k$ for $k \neq j$) of the left hand side in the variable
$z_1$. \ We lose nothing by supposing to start that $z_i \neq z_j
\tmop{for} i \neq j$; this makes all the poles simple. These residues are
Pfaffians that can be inductively evaluated using (\ref{product1}). \ Now
compare these residues with the residues of the right hand side calculated
directly. \ This inductive agreement needs only a check of the case $N = 2$ to
be complete. \ Thus the right and left sides of (\ref{product1}) are rational
functions with equal residues at their simple poles. \ This implies that the
difference of the two sides is a polynomial in $z_1$. \ It is clear that both
sides have finite limits as $z_1 \rightarrow \infty$ and so this polynomial
must be a constant. \ Since both sides are 0 at $z_1 = z_2$ they must be the
same for all $z_1$. \ Anytime $z_i = z_j$ for $i \neq j$, both sides of
(\ref{product1}) are 0 so the equality is true quite generally. \

We are now prepared to state the principal theorem of this paper.

\begin{theorem}
  \label{th:4}Write $\tmmathbf{s}' = (s'_1, \ldots, s'_m)$ with $s_j' \in
  \Sigma_P'$ and $\tmmathbf{s}= (s_1, \ldots, s_n)$ with $s_j \in \Sigma_A'$.
  Recall that for $\#\tmmathbf{s}'$=m and $\#\tmmathbf{s}= n$ both even,
  \[ e_P^+ (\tmmathbf{s}') = e_P^+ (s'_1) \wedge \cdots \wedge e_P^+ (s'_m),
  \]
  and
  \[ e_A^+ (\tmmathbf{s}) = e_P^+ (s_1) \wedge \cdots \wedge e_P^+ (s_n) \]
  are both eigenvectors for the transfer matrix.
  
  The matrix elements of the periodic scaling limit of the Ising spin operator
  $\sigma = \sigma_L$ normalized so that $\langle 0_P, \sigma 0_A \rangle = 1$
  and restricted to a map from $\tmop{Alt}_{\tmop{even}} (W_A^+)$ to
  $\tmop{Alt}_{\tmop{even}} (W_P^+)$ in the basis of eigenvectors $e_P^+
  (\tmmathbf{s}')$ and $e_A^+ (\tmmathbf{s})$ for the transfer matrix is given
  by,
  \[ \langle e_P^+ (\tmmathbf{s}'), \sigma e_A^+ (\tmmathbf{s}) \rangle =
     \Lambda_P (\tmmathbf{s}') \prod^{m + n}_{i < j} \bignone \frac{e^{t_i} -
     e^{t_j}}{e^{t_i} + e^{t_j}} \Lambda_A (\tmmathbf{s}) \]
  where,
  \[ \Lambda_P (\tmmathbf{s}') = e^{- i m \pi / 4} \prod_{j = 1}^m \bignone
     \frac{e^{- i m L \tmop{sh} s'_j}}{\sqrt{2 m L \tmop{ch} s'_j}} e^{- s'_j
     / 2} \lambda_P (s'_j + i \pi / 2), \]
  \[ \Lambda_A (\tmmathbf{s}) = e^{i n \pi / 4} \prod_{j = 1}^n \bignone
     \frac{e^{i m L \tmop{sh} s_j}}{\sqrt{2 m L \tmop{ch} s_j}} e^{- s_j / 2}
     \lambda_A (s_j - i \pi / 2), \]
  and
\end{theorem}
\[ t_j = s'_{j -} = s'_j - i \pi / 2, \tmop{for} j = 1, \ldots, m, \]
\[ t_{j + m} = s_{j +} = s_j + i \pi / 2, \tmop{for} j = 1, \ldots, n. \]
These matrix elements can be used to calculate correlation functions of the
spin operators in the scaling limit. \ For the convergence of the resulting
sums to be manifest, however, it is helpful \ if some non-zero power of the
transfer matrix appears between any two spin operators. \ We will sketch the
issues involved for the scaled two point function on the cylinder.

Return to the lattice formalism on a finite lattice. \ The two point function
on a $(2 \ell + 1) \times (2 m + 1)$ \ doubly periodic lattice is the ratio of
traces,
\[ \langle \sigma_{p, q} \sigma_{0, 0} \rangle = \frac{\tmop{Tr} (\sigma_p V^q
   \sigma_0 V^{n_{}})}{\tmop{Tr} (V^{2 m + 1})}, \]
where $q + n_{} = 2 m + 1$. \ Sending $m \rightarrow \infty$, fixing $p$ and
$q$ one obtains the correlations in the cylindrical limit. \ In this limit the
trace in the denominator is asymptotic to $\lambda_A^{2 m + 1}$ where
$\lambda_A$ is the largest eigenvalue of the transfer matrix. \ We briefly
resurrect the notation from (\ref{eq5}); don't confuse this with the notation
$\lambda_A (s)$ for the solution to the factorization problem. In the limit $n
\rightarrow \infty$ the product $\lambda_A^{- n} V^n$ tends to the orthogonal
projection onto the eigenvector, $0_A,$ associated with the largest eigenvalue
of $V$. \ Thus in the cylindrical limit the correlation tends to,
\[ \langle \sigma_{p, q} \sigma_{0, 0} \rangle = \lambda_A^{- q} \langle 0_A,
   \sigma_p V^q \sigma_0 0_A \rangle . \]
If we introduce a normalized transfer matrix $\hat{V}_P$ (in the periodic
sector) so that $V_P = \lambda_P \hat{V}_P$ (and the eigenvalue for
$\hat{V}_P$ associated with $0_P$ is 1) then the cylindrical correlation
becomes,
\begin{equation}
  \langle \sigma_{p, q} \sigma_{0, 0} \rangle = (\lambda_P / \lambda_A)^q
  \langle 0_A, \sigma_p \hat{V}^q_P \sigma_0 0_A \rangle . \label{cylind}
\end{equation}
On the finite lattice the spin operator, $\sigma_0$, is unitary and
$\sigma_0^2 = 1$. \ It follows that as long as we work in orthonormal bases
for $\tmop{Alt}_{\tmop{even}} (W_A^+)$ and $\tmop{Alt}_{\tmop{even}} (W_P^+)$
the matrix of $\sigma_0$ thought of as a map from $\tmop{Alt}_{\tmop{even}}
(W_P^+)$ to $\tmop{Alt}_{\tmop{even}} (W_A^+)$ is the hermitian conjugate of
the matrix of $\sigma_0$ thought of as a map from $\tmop{Alt}_{\tmop{even}}
(W_A^+)$ to $\tmop{Alt}_{\tmop{even}} (W_P^+)$. \ Replacing $\sigma_p$ by
$\hat{\sigma}_p = \sigma_p / \langle 0_A, \sigma_p 0_P \rangle$ and $\sigma_0$
by $\hat{\sigma}_0 = \sigma_0 / \langle 0_P, \sigma_0 0_A \rangle$ in
preparation for scaling, it is straightforward to understand that, apart from
space translations, the matrices for $\hat{\sigma}_p$ and $\hat{\sigma}_0$ are
hermitian conjugates (observe that, modulo the convergence of the scaling
limit, this shows that it is enough to understand the matrix of the spin
operator from $\tmop{Alt}_{\tmop{even}} (W_A^+)$ to $\tmop{Alt}_{\tmop{even}}
(W_P^+)$ which is what we've focused on in this paper.) We can now use
(\ref{cylind}) to informally understand what happens in the scaling limit to
$\langle \hat{\sigma}_{p, q} \hat{\sigma}_{0, 0} \rangle$. Make the
substitutions $q \leftarrow q m_2^{- 1}$ and $p \leftarrow p m_1^{- 1}$. \
Note that in the limit we are interested in,
\[ (\lambda_P / \lambda_A)^{q m_2^{- 1}} \rightarrow e^{- \frac{\pi q}{4 L}} .
\]
Thus at least informally,
\[ \langle \hat{\sigma}_{p m_1^{- 1}, q m_2^{- 1}} \hat{\sigma}_{0, 0} \rangle
   \rightarrow e^{- \frac{\pi q}{4 L}} \sum_{\tmmathbf{s} \in \Sigma_P'} z
   (\tmmathbf{s})^p \lambda (\tmmathbf{s})^q | \langle e_P^+ (\tmmathbf{s}),
   \hat{\sigma} 0_A \rangle |^2 \bignone, \]
where the sum is over all $\tmmathbf{s}= (s_1, s_2, \ldots, s_n)$ with $s_j
\in \Sigma_P'$ and $s_i \neq s_j$ for $i \neq j$ (including
$\tmmathbf{s}= \emptyset$). Space translation and vertical transfer are given
by,
\[ z (\tmmathbf{s}) = \prod_{j = 1}^n \exp (i m \tmop{sh} (s_j)), \]
\[ \lambda (\tmmathbf{s}) = \prod_{j = 1}^n \exp (- m \tmop{ch} (s_j))
   \bignone . \]

The phase factors $z (\tmmathbf{s})^p$ are the only surviving ones in the
calculation and one can write,
\[ | \langle e_P^+ (\tmmathbf{s}), \hat{\sigma} 0_A \rangle |^2 = \left| \prod_{j =
   1}^n \bignone \frac{e^{- s_j / 2} \lambda_P (s_j + i \pi / 2)}{\sqrt{2 m L
   \tmop{ch} s_j}} \prod^n_{i < j} \bignone \frac{e^{t_i} - e^{t_j}}{e^{t_i} +
   e^{t_j}} \right|^2 \]
Results of this sort were announced in {\cite{FZ03}} for the scaling functions
and derived in {\cite{B01}} for correlations on the finite periodic lattice.

\end{document}